\documentclass[aps,prb,twocolumn,superscriptaddress,usenames,dvipsnames,floatfix,10pt]{revtex4-2}
\usepackage{bm}
\usepackage[T1]{fontenc}
\usepackage{amsmath}
\usepackage{graphicx}
\usepackage[dvipsnames]{xcolor}
\usepackage{tabularx}
\usepackage{multirow}
\usepackage{colortbl}
\usepackage{soul}
\usepackage{cancel}
\usepackage{textcomp}
\usepackage{gensymb}

\usepackage{hyperref}
\hypersetup{colorlinks=true, linkcolor=blue, citecolor=blue, urlcolor=blue}

\usepackage{xr}
\usepackage[normalem]{ulem}
\hypersetup{colorlinks=true, linkcolor=blue, citecolor=red, urlcolor=magenta, pdftitle={Large anti-magnetoelectricity from spin and orbitals in ferroelectric BiCoO3}, pdfauthor={M. Braun, B. Guster, E. Bousquet}}\usepackage{natbib}
\definecolor{gray}{rgb}{0.8,0.8,0.8}

\def\s2s21  {{$\sqrt{2}\times\sqrt{2}\times1$}}

\begin{document}



\title{First-principles study of KCoF$_3$: Jahn-Teller effect, dynamical magnetic charges, magnetoelectric multipoles and antimagnetoelectricity}

\author{Bogdan Guster}
\affiliation{Physique Th\'eorique des Mat\'eriaux, Q-MAT, Universit\'e de Li\`ege, B-4000 Sart-Tilman, Belgium}
\affiliation{European Theoretical Spectroscopy Facility, www.etsf.eu}
\author{Maxime Braun}
\affiliation{Physique Th\'eorique des Mat\'eriaux, Q-MAT, Universit\'e de Li\`ege, B-4000 Sart-Tilman, Belgium}
\affiliation{Univ. Lille, CNRS, Centrale Lille, ENSCL, Univ. Artois, UMR 8181-UCCS-Unité de Catalyse et Chimie du Solide, F-59000 Lille, France}
\author{Eric Bousquet}
\affiliation{Physique Th\'eorique des Mat\'eriaux, Q-MAT, Universit\'e de Li\`ege, B-4000 Sart-Tilman, Belgium}

\begin{abstract}
We study from \textit{ab~initio} density functional theory calculations the structural and magnetic properties of the crystal KCoF$_3$.
We found that the experimentally reported cubic to tetragonal phase transition is due to an electronic first-order Jahn-Teller effect from the R zone boundary point.
We also obtain that the magnetic ground state is the G-type antiferromagnetic order, in agreement with the R-point Jahn-Teller distortion and that the magnetic moment of the Co atoms contains a strong orbital contribution ($m_L=0.95$ $\mu_B$ in the cubic phase and 0.55 $\mu_B$ in the tetragonal phase).
Furthermore, we compute the dynamical magnetic effective charges and show that it is zero by symmetry for the Co and they can reach a value as large as 200 $10^{-2}\mu_{\text{B}}/\text{\r{A}}$ for the apical F anion.
This large magnetic effective charge comes from the spin-orbit coupling (50\% of the response is from the orbital moment) contrary to the rare-earth manganites and ferrites with similar order of magnitude but originating from the exchange striction mechanism.
The fact that the dynamical magnetic effective charges are non-zero also proves that the tetragonal phase of KCoF$_3$ is antimagnetoelectric with a large magnetic sublattice magnetoelectric response of 210 ps/m per spin-channel.
We also discuss the generality of these magnetic effective charges.

\end{abstract}

\maketitle

\section{Introduction}

KMF$_3$ fluoride perovskite crystals (with M a 3d transition metal such as Fe, Cu, Mn, Co, Cr, Ag) have attracted significant interest from a fundamental physics and materials science perspective due to their unique properties (first-order Jahn-Teller effect, orbital ordering, Mott insulators, metal-insulator transition, electron localization, low-dimentional magnetism, magnetoelasticity, etc)~\cite{gnezdilov2012,lee2012,zhou2011,Hirakawa1960,Oleaga2015, hirakawa1970, Koteras2025, li2021, clark2016, iglamov2014, legut2013, yuan2012}.
They provide model systems for exploring strongly localized electron physics, low dimentional magnetism, or lattice dynamics within a simple symmetric cubic framework, making them ideal for testing theoretical models~\cite{Varignon2019,li2021,pavarini2008,leonov2008,leonov2010}.
Among them, the case of KCoF$_3$ has been less studied and those studies focused mainly on its magnetic properties, while its electronic, structural or vibrational properties were scarce.
For example, a structural phase transition has been observed experimentally by the observation of cubic to tetragonal cell parameters change at the critical temperature of the antiferromagnetic phase transition ($T_N=115 K$~\cite{Okazaki1961,Hirakawa1960,Oleaga2015}) but no further insight was reported regarding the space group and the atomic positions or on the exact origin of this structural transition (hypothetically reported as an elastic Jahn-Teller effect~\cite{Hirakawa1960,Okazaki1961,Okazaki1959}).

In this article, we focus on KCoF$_3$ by doing density functional theory (DFT) calculations of the structural, electronic, magnetic and vibrational properties of the cubic and tetragonal phase of KCoF$_3$.
This allows us to clarify the first-order Jahn-Teller origin of the phase transition and that the tetragonal phase is the ground state of the system. 
We also scrutinized the magnetism and confirmed the G-type AFM ordering observed experimentally and found that the spin-orbit coupling (SOC) is important, which gives a non-negligible orbital contribution to the magnetic moment of the Co atom.
Going beyond those classical studies, we originally computed the dynamical magnetic effective charges~\cite{Iniguez2008,Braun2024} and the atom magnetoelectric multipoles of KCoF$_3$ and found that they are surprisingly large for a ``non-functional'' material and that the spin and orbital contributions are on equal footing.
Those later calculations allows us to conclude that KCoF$_3$ is anti-magnetoelectric~\cite{Verbeek2023}.

The article is organized as follows: in Sec. \ref{sec:comp_det} we expand on the used computational framework.
In Sec. \ref{sec:results}, we first report on the magnetic ground state, followed by the associated lattice distortion resulting from a Jahn-Teller instability.
We then continue by calculating the dynamical magnetic charges (DMC) and compare their amplitude within the existing literature, and calculate the spin-channel dependent anti-magnetoelectric response from both spin and orbit contributions. A multipolar decomposition of the DMC is also done.
We conclude in Sec. \ref{sec:Conclusions}. 
In Appendix \ref{annex:DMC} we remind of the theoretical background pertaining to (anti-)magnetoelectricity and associated magnetoelectric multipoles.

\section{Computational details}
\label{sec:comp_det}

Our density functional theory (DFT)  calculations were performed with the ABINIT code version 9.10.5 \cite{Gonze2020} in the PBEsol+U framework~\cite{Torrent2008} for the exchange-correlation functional.
We used the JTH projected augmented wave atomic potentials (v1.1)~\cite{Jollet2014} and we included the spin-orbit coupling (SOC) treatment (unless stated otherwise).  
A Hubbard-U correction of 4.0 eV was applied to the 3d Co shell, adjusted to match the total (spin + orbital) experimental magnetic moment of 3.33 $\mu_B$~\cite{Scatturin1961}.
We have checked that the qualitative results are not affected by the U value (see Supplementary Material Tab. S.I.).
The Kohn-Sham pseudo-wave functions were expanded over a plane-wave basis set with a kinetic energy cut-off of 24 Ha and with a cut-off for the PAW double grid kinetic energy cut-off of 48 Ha.
We used a strict convergence threshold of $2.7\times10^{-12}$ eV for the total energy on the self-consistent cycle, to determine the DMCs accurately. 
The BZ of the magnetic unit cell was sampled with a Monkhorst-Pack k-point grid of 6x6x4 points in the G-AFM unit cell ($\sqrt{2}\times\sqrt{2}\times 2$ with respect to the 5-atoms cubic unit cell). 
We also used the density functional theory perturbation (DFPT) to determine the eigenvalues and eigendisplacements of phonons as well as the Born effective charges~\cite{Gonze1997}, that are key ingredients to calculate the lattice contribution to the anti-magnetoelectric response.
The orbital and spin magnetization associated with a specific atom $I$ is computed by integrating the local magnetization density $m_\alpha(\vec{r})$ within a defined sphere $\Omega_I$ centered at the atom's position $\vec{R}_I$ ~\cite{Romero2020}, and 
the radius of the sphere is  equal to the one of the PAW atom dataset~\cite{Romero2020}. 
This muffin-tin approximation is precise enough for the localized d-orbitals of Co for both spin and orbital magnetic moment~\cite{Zhang2019} and with a very good CPU time consuming/precision with respect to a more precise implementation, e.g. Zwanziger et al.~\cite{Zwanziger2023}.
The magnon dispersion and the calculations of superexchange parameters were obtained through the TB2J code version 0.6.6~\cite{Xu2021}.

Finally, we determined the DMC by using a finite differences method where we displace each atom along the 3 Cartesian axes with the following amplitudes: $\pm$0.01\r{A},$\pm$0.02\r{A},$\pm$0.03\r{A}~\cite{Braun2024}.

\section{Results}
\label{sec:results}

\begin{figure}
    \centering
    \includegraphics[width=1\linewidth]{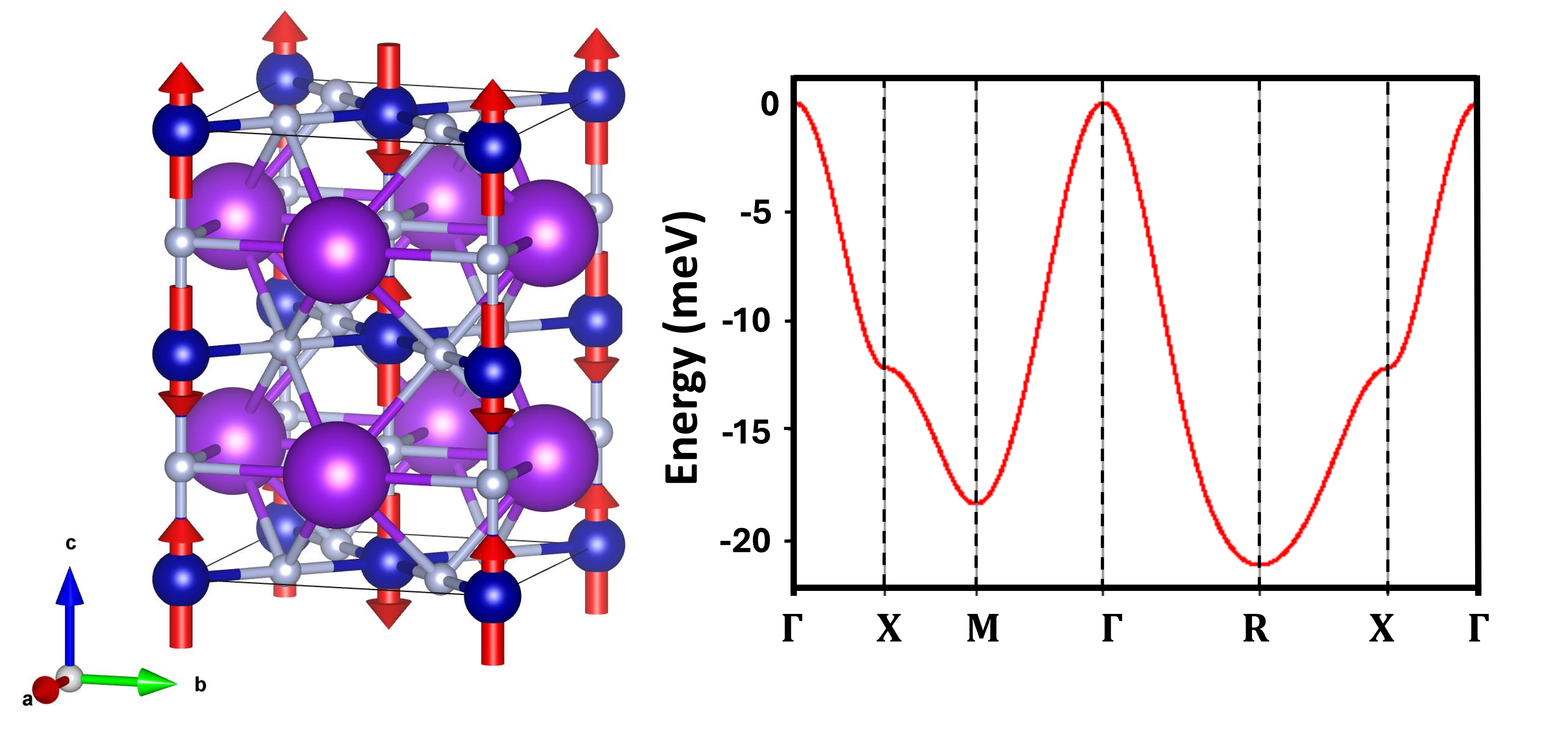}
    \caption{(Left) Magnetic structure of KCoF$_3$ showing the ground state G-type antiferromagnetic ordering. The purple spheres represent K atoms, blue spheres represent Co atoms with alternating spin directions (indicated by red vectors), and light gray spheres represent F atoms forming octahedra around Co sites. The crystallographic axes are indicated in the bottom left corner. (Right) Calculated magnon dispersion along high-symmetry paths in the FM cubic Brillouin zone. The dispersion shows characteristic energy minima at the R point corresponding to the configuration shown  in left panel.  $\Gamma$= (0, 0, 0), X= (1/2, 0, 0), M = (1/2, 1/2, 0) and R= (1/2, 1/2, 1/2) in units of the cubic reciprocal lattice vectors. }
    \label{MagStructure}
\end{figure}

\subsection{Cubic phase}
Above 115 K, KCoF$_3$ is found in a non magnetic cubic perovskite phase described by the space group $Pm
\overline{3}m$ (\#221) with 5 atoms in the unit cell. 
Below this critical temperature $T_{N} = 115~K$~\cite{Okazaki1961,Hirakawa1960,Oleaga2015}, KCoF$_3$ is antiferromagnetic (AFM) with an associated cubic propagation wavevector $k= (1/2, 1/2, 1/2)$, which corresponds to the so called G-type AFM magnetic ground state (Fig.\ref{MagStructure}), i.e. each Co magnetic moment is surrounded by Co first nearest neighbours with opposite magnetic moments. 

In our 0 K DFT calculations, we found that the G-AFM order is the lowest energy magnetic phase in the cubic phase as obtained through both total energy calculations (see Tab.~\ref{tab:cubic_phases}) and the magnon dispersion (see Fig. \ref{MagStructure}  where the lowest energy point in the dispersion is at the $k= (1/2, 1/2, 1/2)$ point corresponding to the G-AFM ordering).
The calculated lattice parameters (Tab.~\ref{tab:cubic_phases}) are in good agreement with the experimental values of 4.069 $\pm$ 0.001 \AA~\cite{Okazaki1961,Wang2013}.
As our calculations were done within non-collinear DFT regime and including the spin-orbit coupling, we found that all the magnetic phases break the cubic structure with a slight tetragonal distortion of the cell (see Tab. \ref{tab:cubic_phases}) but keeping all atoms in their high symmetry positions. 
This broken cubic cell parameter by the AFM orders is supported by the corresponding magnetic space group symmetry as all of the AFM orders give a tetragonal magnetic space group when considering the magnetic moment direction (i.e. the magnetic moments are treated as pseudovectors and not as simple scalar values as done in collinear calculations, see Tab.~\ref{tab:cubic_phases}).
We also stress that none of these AFM-related magnetic space groups allows for magnetoelectricity.

\begin{table}[!hptb]
\caption{Relative energy with respect to the FM phase and lattice parameters of the common commensurate magnetic phases of the high symmetry KCoF$_3$,  along with the associated magnetic space groups (magnetic moment along the z direction). 
We can see that the magnetic ordering breaks the cubic symmetry when considering magnetic moment orientation within the non-collinear regime calculation and the inclusion of the spin-orbit coupling. 
}
\begin{tabular}{ccccc}
\hline\hline
\multicolumn{1}{c|}{\begin{tabular}[c]{@{}c@{}}Magnetic\\ Phase\end{tabular}} & \multicolumn{1}{c|}{$\Delta E$ (meV/f.u.)} & \multicolumn{1}{c|}{a}     & \multicolumn{1}{c|}{c}     & Mag S.G. \\ \hline
\multicolumn{1}{c|}{FM}             & \multicolumn{1}{c|}{  0}     & \multicolumn{1}{c|}{4.077} & \multicolumn{1}{c|}{4.012} &  $P4/mm'm'$         \\ \hline
\multicolumn{1}{c|}{A-AFM}          & \multicolumn{1}{c|}{-32}     & \multicolumn{1}{c|}{4.079} & \multicolumn{1}{c|}{4.002} &  $P_P4'/mmm'$         \\ \hline
\multicolumn{1}{c|}{C-AFM}          & \multicolumn{1}{c|}{-51}     & \multicolumn{1}{c|}{4.069} & \multicolumn{1}{c|}{4.010} &  $P_2c4/mm'm'$        \\ \hline
\multicolumn{1}{c|}{G-AFM}          & \multicolumn{1}{c|}{-79}     & \multicolumn{1}{c|}{4.071} & \multicolumn{1}{c|}{4.001} &  $P_I4/mm'm'$        \\ \hline
\end{tabular}
\label{tab:cubic_phases}
\end{table}

Our calculated total magnetic moment of the Co atoms is found to be 3.43 $\mu_{B}$, which is in good agreement with the experimental value of 3.33 $\mu_{B}$~\cite{Scatturin1961} and we found it to be the same value irrespective of the magnetic ordering.
This total calculated magnetic moment decomposes into a spin component of 2.50 $\mu_{B}$ and an orbital contribution of 0.93 $\mu_{B}$. 
The free Co$^{2+}$ ion with $d^{7}$ configuration can theoretically host a magnetic moment of 3 $\mu_{B}$ for both spin and orbital contributions, totalizing 6 $\mu_{B}$ in spherical symmetry.
This ideal value can be reduced through several effects: (i) the local rotational symmetry associated with the quenching of the orbitals can greatly reduce the free ion maximum value~\cite{Koo2020},
(ii) 
the spin-orbit coupling allows for a magnetic moment transfer from the spin to the unquenched orbital magnetic moment,
(iii) 
the magnon-phonon interaction itself is further enhanced owing to the unquenched orbital magnetic moment \cite{Isu1978} and a residual orbital magnetic moment can be present in the cubic crystal field~\cite{Kanamori1957}.
The orbital moment value we obtain in KCoF$_3$ is in line with the equivalent ones in octahedral coordination, e.g. CoO$_4$X$_2$ (X=Cl, Br, S, Se)~\cite{Koo2020,Whangbo2022}. 
Although the highest reported value of Co$^{2+}$ to our knowledge is found in SrTi$_{0.88}$Co$_{0.04}$Nb$_{0.08}$O$_3$ where the total magnetic moment is 5.95 $\mu_{B}$, i.e. close to the free ion value~\cite{Oey2020}.
Nonetheless, the orbital magnetic moment in KCoF$_3$ is  larger than in the prototypical transition-metals of  VI$_3$ (0.59 $\mu_{B}$)~\cite{Hovancik2023}, Fe$_3$O$_4$ (0.44 $\mu_{B}$) or CrO$_2$ (0.06 $\mu_{B}$)~\cite{Huang2004a}.
The presence of an important orbital contribution to the magnetic moment can have an impact on, e.g., the magnon dispersion~\cite{neuman2020} as reported in a compound similar to KCoF$_3$, RbCoF$_3$~\cite{nouet1973}.

According to Dubrovin et al.~\cite{Dubrovin2021}, the calculation of the phonon dispersion shows that there is no unstable mode in the cubic phase (see Supplementary Material Fig. S.2), which means that the structure is found to be stable at the harmonic level.
However, this does not explain the cubic to tetragonal transition reported experimentally.

\subsection{Tetragonal phase}

As we have writen above, experimentally  below the critical temperature of 115 K a lattice distortion driving KCoF$_3$ from a cubic phase to a tetragonal phase is simultaneous to the appearance of the magnetic order~\cite{Okazaki1959,Okazaki1961}, similarly to KMnF$_3$ but for which it has been proved that the coincidence of the structural transition and the Néel temperature is essentially accidental~\cite{carpenter2012}.
However, no experimental report on the space group or atom positions below 115 K is reported, only the values of the cell parameters and their tetragonal splitting are reported~\cite{Okazaki1959,Okazaki1961}. 
Some authors mention that the transition can be purely elastic \cite{Kanamori1960,Scott1974}, but this is based on a more speculative theoretical analysis and it is not supported by any experimental work or theoretical calculations.

\begin{figure}[!hptb]
    \centering
    \includegraphics[width=0.95\linewidth]{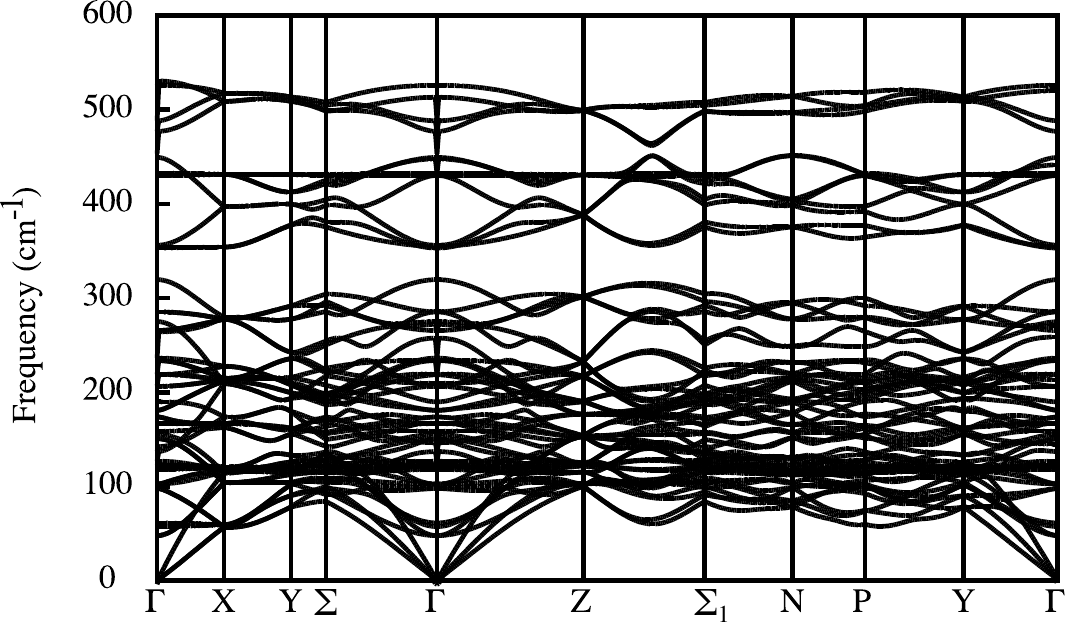}
    \caption{Phonon dispersion curves of KCoF$_3$ in the calculated (collinear DFT regime) ground state G-AFM $I4/mcm$ phase. 
    $\Gamma$ = (0, 0, 0), X = ( 0, 0, 1/2), Y = ( -0.26, 0.26, 1/2), $\Sigma$ = ( -0.38, 0.38, 0.38), Z = ( 1/2, 1/2, -1/2), $\Sigma_1$ = (0.38, 0.62, -0.38), N = ( 0, 1/2, 0), and P = (1/4, 1/4, 1/4) in units of the cubic reciprocal lattice vectors.}
    \label{fig:phonon_disp}
\end{figure}

\begin{figure}
    \centering
    \includegraphics[width=1\linewidth]{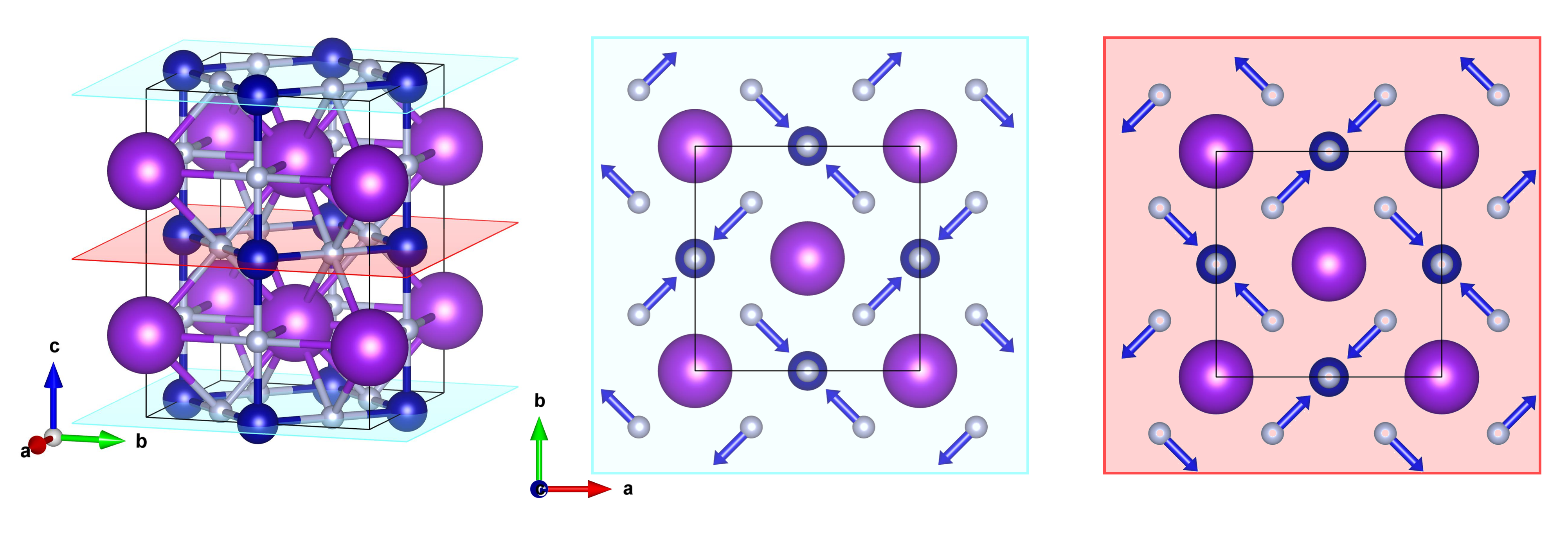}
    \caption{Visualization of the atomic displacements associated with the $R_3^-$ Jahn-Teller mode. 
    (Left) Three-dimensional perspective of the crystal structure with two consecutive cobalt planes highlighted in cyan and red. (Center, Right) Top-view projections of the cobalt planes  illustrating the in-plane atomic motion patterns. }
    \label{fig:VisualisationJTMode}
\end{figure}

High spin Co d$^7$ orbitals in the octahedra environment should be Jahn-Teller active as the two electrons of the minority channel break the $t_{2g}$ 3-fold degeneracy of the octahedral cubic crystal field~\cite{Schmitt2020}. 
This situation should correspond to the so-called first-order Jahn-Teller effect~\cite{Kugel1982,Okazaki1961,Bersuker1984}.
Previous first-principles collinear DFT calculations go into this direction, as Dubrovin et al.~\cite{Dubrovin2021} report that the cubic phase in its G-AFM state has no unstable phonon in the whole Brillouin zone (Fig. 5 of Ref.~\cite{Dubrovin2021}) and Varignon et al.~\cite{Varignon2019} report that the tetragonal phase of KCoF$_3$ comes from a first order Jahn-Teller instability, which is associated with the cubic $R$ zone boundary point pattern of distortion.
As the first-order Jahn-Teller effect appears at the first order in an energy expansion, it is not seen through unstable phonons as the phonons correspond to a second order energy derivatives.
We confirm those results in our collinear calculations as we observe a sudden jump in energy  once the cubic symmetry is broken through Jahn-Teller pattern of distortion (see Supplementary Material Fig. S.1), and thus before any lattice distortion is allowed, which is a fingerprint of an electronic instability of the first order Jahn-Teller type~\cite{Varignon2019}.
The full relaxation of the cell gives the typical Jahn-Teller distortion of the R zone boundary point (R$_3^-$ label with origin on the K atom~\cite{Schmitt2020}) with the tetragonal space group $I4/mcm$ illustrated in Fig.~\ref{fig:VisualisationJTMode}.
The calculation of phonon dispersion in this relaxed lower energy phase (see Fig.~\ref{fig:phonon_disp}) does not exhibit any unstable mode and if we break all symmetry operations and fully relax the crystal, the system stays in this $I4/mcm$ phase, such that we can conclude that this phase is locally stable.

\begin{figure}
    \centering
    \includegraphics[width=0.99\linewidth]{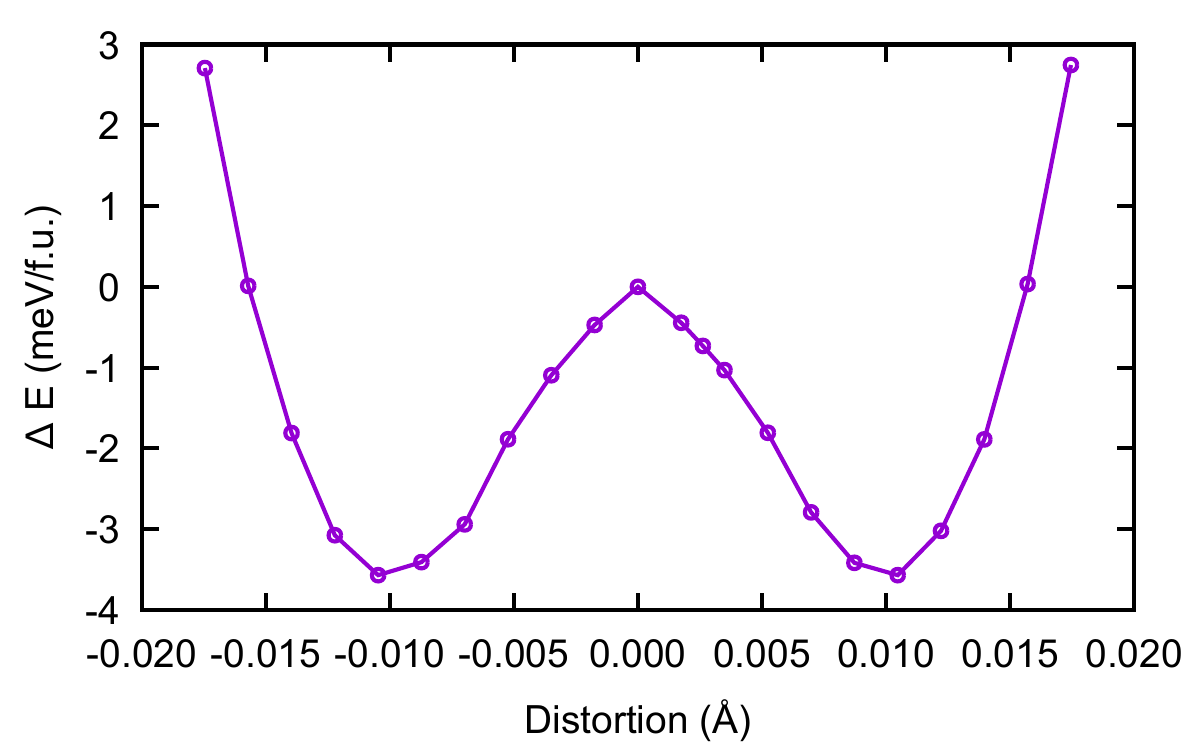}
    \caption{Energy evolution under the first-order type Jahn-Teller rigid distortion in KCoF$_3$ in the G-AFM magnetic state including SOC but at fixed cell parameter (the one of the high symmetry reference, see Table S.I). The zero energy is taken from the $P_I4/mm'm'$ high symmetry phase including SOC (Table S.I). Under the full cell relaxation, a total gain of 23.1 meV/f.u. is obtained, showing that the elastic contribution is important as at frozen cell parameter as shown in this figure the gain of energy is a bit less than 4 meV/f.u.}
    \label{fig:delta_e}
\end{figure}

In our search for local minima and including the spin-orbit coupling (non-collinear DFT calculation regime), we found   
(i) the $I4/mcm$ phase with a gain of energy of 23.1 meV/f.u. (with respect to the G-AFM high symmetry reference with spin-orbit coupling, see Fig.~\ref{fig:delta_e}) 
(ii) a purely elastic distorted phase, with the space group $P2_1/c$, which reduces the energy by 0.7 meV/f.u., hence a gain of energy much smaller than the $I4/mcm$ phase, and 
(iii) another Jahn-Teller distortion phase associated with a M-point distortion (M$_3^+$ label giving the $I4/mbm$ space group~\cite{Schmitt2020}) and lowering the energy by 11.1 meV/f.u., again higher than the $I4/mcm$ phase.
Hence, according to our calculations, the G-AFM $I4/mcm$ phase is  the ground state of KCoF$_3$ and it comes from a first-order Jahn-Teller electronic instability combined with a lattice distortion associated with the R-point  distortion~\cite{Schmitt2020}.

In this ground state phase, we found that the magnetic crystal space group is $I4'/mcm'$, i.e. with a magnetic easy-axis along the $c$-axis lattice vector direction, in agreement with experimental reports~\cite{Oleaga2015}. 
This magnetic symmetry allows for the inverse piezomagnetic effect (tensor $\wedge^T$ coupling magnetization direction to strain). 
Consequently, the total free energy cannot be easily separated into intrinsic magnetocrystalline anisotropy and magnetoelastic contributions without complex simultaneous relaxation of lattice and magnetization, which is beyond the scope of the present study on dynamical charges.

\begin{figure}[!hptb]
    \centering
    \includegraphics[width=0.45\textwidth]{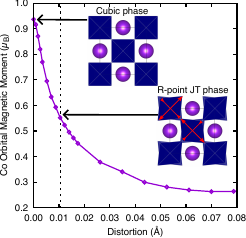}
    \caption{Evolution of Co orbital magnetic moment of KCoF$_3$ with respect to the Jahn-Teller distortion amplitude in it G-AFM magnetic ground state. 
    The relaxed Jahn-Teller distortion amplitude (0.01 \r{A}) is reported by the vertical dashed line.    }\label{fig:omm_gafm}
\end{figure}

The only internal degree of freedom falls on the basal F atom at the 8h Wyckoff site ( \textit{x} , \textit{x}+$\frac{1}{2}$, 0), with \textit{x}=0.03 \r{A}. 
The distortion corresponds to the typical Jahn-Teller distortion from the R-point~\cite{Schmitt2020}, i.e. an in-plane stretching of the octahedra in anti-phase with both in- and out-of-plane neighbouring octahedra (see inset of Fig.~\ref{fig:omm_gafm}).
We would like to mention here that the cubic phase cannot stay in cubic symmetries with the G-AFM magnetic phase while including the spin-orbit coupling.
Indeed, the magnetic space group is the tetragonal P$_I$4/mm'm' (for a  magnetic easy-axis along c ~\cite{Oleaga2015}) such that the electronic instability is always relaxed. 
In this situation with the spin-orbit coupling on, we cannot probe the sudden jump in energy due to the electronic instability~\cite{Varignon2019} (see Fig.~\ref{fig:delta_e}) but we can recover it when doing the collinear calculations (see Fig.S.1).
The energy well is closer to a smooth double well, but we can still observe a dominant linear evolution of the energy around the zero distortion as if two single wells would be present~\cite{Varignon2019}.

Let us address the implication of the Jahn-Teller electronic instability on the magnetism of KCoF$_3$. 
In Fig.~\ref{fig:omm_gafm} we show the evolution of the Co atom orbital angular momentum versus the Jahn-Teller distortion. 
The orbital magnetic moment of Co atoms has a large value in the cubic phase of about 0.95 $\mu_B$.
Then, once the Jahn-Teller distortion is frozen into the structure, we find this orbital magnetic moment to first decrease strongly for small distortions (up to roughly 0.03-0.04 \AA) and beyond this range of distortion amplitudes it slowly stabilizes to about 1/3 of the initial value (see Fig. \ref{fig:omm_gafm}). 
We would like to stress that the energy minimum of the $I4'/mcm'$ Jahn-Teller ground state phase (i.e. at around 0.01 \r{A}) is reached much earlier than the orbital magnetization stabilization, giving a value of the orbital magnetic moment of 0.55 $\mu_B$ (vertical dashed line in Fig. \ref{fig:omm_gafm}).
However, within the same regime of distortion amplitude, the spin contribution to the magnetic moment does not vary significantly (0.4\%).

Hence, we can see from this analysis that the orbital magnetic moment is strongly sensitive to the Jahn-Teller pattern of atomic distortions, so that we will scrutinize in the next section  how each individual atom displacement affects the magnetic moment and magnetism of KCoF$_3$.

\subsection{Dynamical magnetic charges}
One way to check how individual atomic motion can affect the magnetism of a crystal is through the dynamical magnetic charges (DMC) (see Annex~\ref{annex:DMC}).
In Table~\ref{tab:dmc_values}, we report our calculated DMC tensor elements for each nonequivalent atom and by decomposing the spin ($\boldsymbol{S}$ column in the table) and orbital ($\boldsymbol{L}$ column in the table)  magnetic moment contributions.
We can first see that the DMC associated with the Co atom is zero, a surprising result considering that one could expect that the magnetic atom would carry the highest DMC values~\cite{Ye2014, Braun2024}.  
However, a symmetry analysis confirms this result as the 4d Wyckoff position where the Co atoms lie, which corresponds to the 
$m.m'm'$ magnetic point group 
, should be strictly zero by symmetry.

\begin{table}[!hptb]
\caption{DMC tensor elements of each inequivalent atoms displayed in the crystal basis in the $I4'/mcm'$ phase of KCoF$_3$ based on the symmetry analysis. The \textit{Symm.} column gives the magnetic point group of the Wyckoff site given in the second column (see also Tab. S.VI) and the fourth column gives the non-zero DMC tensor elements. In the last two columns, the  corresponding calculated values for both spin ($\boldsymbol{S}$) and orbital ($\boldsymbol{L}$) contributions are given in $10^{-2}\mu_B/\textup{\AA}$.The values $<3$ are those lower than the precision of the calculations.}
\begin{tabular}{cccccc}
\hline\hline
\multicolumn{1}{c|}{Atom} & \multicolumn{1}{c|}{Wyc.} & \multicolumn{1}{c|}{Symm.} & \multicolumn{1}{c|}{non-zero components} & \multicolumn{1}{c|}{$\boldsymbol{S}$}        & $\boldsymbol{L}$        \\ \hline
\multicolumn{1}{c|}{K}    & \multicolumn{1}{c|}{4a}      & \multicolumn{1}{c|}{4'22'}    & \multicolumn{1}{c|}{$Z^{*m}_{xx}=-Z^{*m}_{yy}$ }             & \multicolumn{1}{c|}{-54}      & -62      \\ \hline
\multicolumn{1}{c|}{Co}   & \multicolumn{1}{c|}{4d}      & \multicolumn{1}{c|}{m.m'm'}     & \multicolumn{1}{c|}{-}             & \multicolumn{1}{c|}{-}        & -        \\ \hline
\multicolumn{1}{c|}{F$_{api}$} & \multicolumn{1}{c|}{4b}      & \multicolumn{1}{c|}{-4'2m'}   & \multicolumn{1}{c|}{$Z^{*m}_{xx}=Z^{*m}_{yy},Z^{*m}_{zz}$}              & \multicolumn{1}{c|}{90, $<3$ } & 96, -10  \\ \hline
\multicolumn{1}{c|}{F$_{bas}$} & \multicolumn{1}{c|}{8h}      & \multicolumn{1}{c|}{m.2'm'}    & \multicolumn{1}{c|}{$Z^{*m}_{xz}=Z^{*m}_{yz},Z^{*m}_{zx}=Z^{*m}_{zy}$}              & \multicolumn{1}{c|}{30, $<3$ } & 24, $<$ 3 \\ \hline 

\end{tabular}
\label{tab:dmc_values}
\end{table}

The K atom at the 4a Wyckoff position corresponds to the $4'22'$ magnetic point group, which allows for the $xx$ and $yy$ DMC tensor components to be non zero and with $Z^{*m}_{xx}=-Z^{*m}_{yy}$.
This is exactly what we obtain from our DFT calculations where the spin contribution ($-54$ $\times10^{-2}\mu_B/\textup{\AA}$) is smaller than the orbital contribution ($-62$ $\times10^{-2}\mu_B/\textup{\AA}$), and giving a total of $-116$ $\times10^{-2}\mu_B/\textup{\AA}$.

The apical F atom at the 4b Wyckoff site is associated with the $-4'2m'$ magnetic point group, which allows the the diagonal $xx$, $yy$ and $zz$ DMC tensor components to be non zero and where $Z^{*m}_{xx}=Z^{*m}_{yy}\neq Z^{*m}_{zz}$.
Here again, our DFT calculations reproduce those non-zero DMC matrix elements where the $xx/yy$ components are high (90 $\times10^{-2}\mu_B/\textup{\AA}$ and 96 $\times10^{-2}\mu_B/\textup{\AA}$ for the spin and orbital contributions respectively, making a total of 186 $\times10^{-2}\mu_B/\textup{\AA}$) and the $zz$ component comparatively small ($<3$ $\times10^{-2}\mu_B/\textup{\AA}$ for the spin part and -10 $\times10^{-2}\mu_B/\textup{\AA}$ for the orbital part).

At last, the basal fluorine located at the 8h Wyckoff position allows for the $yz$ and $zy$ off-diagonal DMC elements to be non zero and with $Z^{*m}_{xz}=Z^{*m}_{yz}\neq Z^{*m}_{zx}=Z^{*m}_{zy}$.
Our DFT results also find a non-zero and large value for the $yz$ component ($30\times10^{-2}\mu_B/\textup{\AA}$ for the spin part and $24\times 10^{-2}\mu_B/\textup{\AA}$ for the orbital part) and a negligible value for the $zy$ component.

We can first note that the spin and orbital contributions are roughly on par for all atoms, meaning that the orbital magnetization contribution with SOC cannot be neglected for this crystal. 
The second note is that the DMC values of KCoF$_3$ are rather large when comparing with other crystals with 3d transition metal cations.
The largest value obtained is for the F atom at the 4b Wyckoff site (186 $\times10^{-2}\mu_B/\textup{\AA}$), which is about six times larger than the one reported for BiCoO$_3$ (31 $\times10^{-2}\mu_B/\textup{\AA}$)~\cite{Braun2024} and two orders of magnitude higher than in Cr$_2$O$_3$ (6$\times10^{-2}\mu_B/\textup{\AA}$)\cite{Ye2014,Tillack2016}.
This is smaller but of the same order of magnitude than the artificially designed magnetoelectric KITPite compound (CaAlMn$_3$O$_7$~\cite{Ye2014}) with a  calculated value reported of 341 $\times10^{-2}\mu_B/\textup{\AA}$ (see Table~\ref{HighestDMCTable}). 
The largest reported DMC values are for the hexagonal perovskites HoMnO$_3$ and LuFeO$_3$, with a value of 445 and 665 $\times10^{-2}\mu_B/\textup{\AA}$, respectively (Table~\ref{HighestDMCTable}).
Those large values in hexagonal perovskites originate from geometric spin frustration and magnetostriction effects~\cite{Ye2015}, while KCoF$_3$ stands out as an exceptional case because its important DMC values are not from these conventional mechanisms.
The large DMC values in KCoF$_3$ arise from a different pathway involving orbital degrees of freedom and Jahn-Teller distortions, which represent a potential alternative route to explore substantial dynamical magnetic charges in cubic perovskite systems.

\begin{table}[!hptb]
\caption{Non exhaustive table of comparison of previously reported DMC values (in $10^{-2}\mu_B/\textup{\AA}$), for which atom and for which tensor element (the largest over all other atoms is reported). Beside the reference magnetoelectric crystal Cr$_2$O$_3$, we report the largest cases we could find in the literature. Indexes $api$ and $bas$ refer to the apical, respectively basal positions in their respective structures.}
\begin{tabular}{l|c|c|c|c}
\hline\hline
Material                                 & Atom      & Element & \multicolumn{1}{c|}{largest DMC value} & Reference\\ \hline
Cr$_2$O$_3$     & Cr        & xy      & 6          &   \cite{Ye2014,Tillack2016}            \\ 
BiCoO$_3$              & Co        & xx      & 31     &      \cite{Braun2024}               \\ 
KCoF$_3$                                 & F$_{api}$ & xx      & 186    & present                  \\ 
CaAlMn$_3$O$_7$             & Mn        & xx      & 341           & \cite{Ye2014}           \\ 
HoMnO$_3$                   & O$_{bas}$ & xx      & 445        & \cite{Ye2015}              \\ 
LuFeO$_3$                & O$_{bas}$ & xx      & 665     & \cite{Ye2015}                    \\ \hline
\end{tabular}
\label{HighestDMCTable}
\end{table}

\begin{figure}
    \centering
    \includegraphics[width=.8\linewidth]{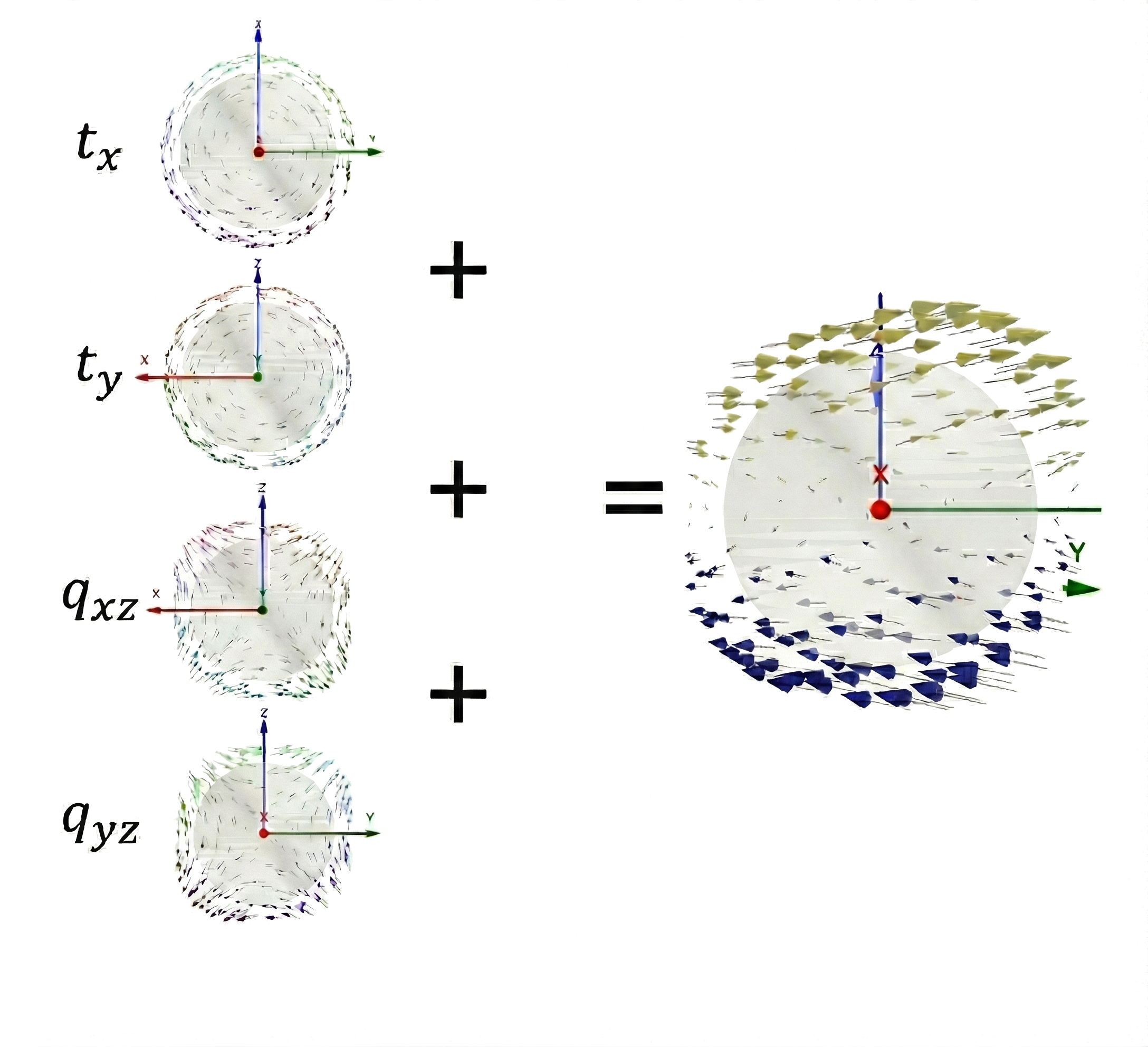}
    \caption{Schematic representation of the DMC decomposition into multipoles components of the F$_\text{base}$ at the 8h Wyckoff site. See Appendix~\ref{Appendix:Multipoles} for the $t$ and $q$ multipole notations.}
    \label{fig:F2DmcDecomposition}
\end{figure}

\begin{figure}
    \centering
    \includegraphics[width=0.722\linewidth]{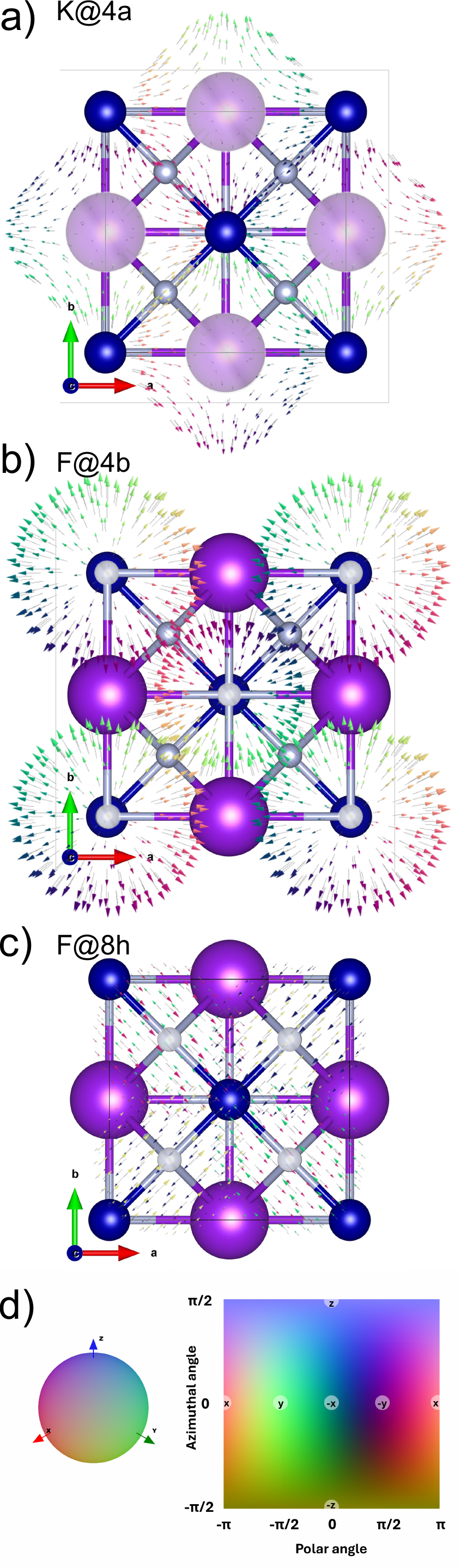}
    \caption{Visualization of the different DMC vectorial representation of  (a) the K atoms at the 4a Wyckoff position, (b) the F atoms at the 4b Wyckoff position, respectivelly and (c) the F atoms at the 8h Wyckoff positions. The associated colormap representation is show in panel (d).}
    \label{fig:dmc_visual}
\end{figure}

To help visualizing the DMC's, we represent the DMC site tensor of each Wyckoff position using a vector field approach (see Fig.~\ref{fig:dmc_visual}a), b) and c))~\cite{Braun2024}.
This vector field is defined by the linear transformation $\mathbf{y} = \mathbf{A}\mathbf{x}$, where $\mathbf{y}$ denotes the resulting vector field, $\mathbf{A}$ is the DMC tensor acting as the Jacobian matrix, and $\mathbf{x}$ represents the atomic displacement vector. 
The DMC tensor, serving as the linear operator, maps displacements to corresponding changes in magnetization, thereby providing a complete characterization of the site-specific magnetoelectric coupling (moving a single atom is polar by construction in ionic crystals and if this atom motion is associated to a change in magnetization, it works as a magnetoelectric response). 
Due to the reciprocal nature of magnetoelectricity, this tensor representation can also be interpreted in the inverse sense: it describes the atomic displacement induced by an external magnetic field applied in a given direction. 
Specifically, if a magnetic field is applied along the x-axis, the corresponding atomic displacement response can be determined by examining the direction of the associated vector field arrows along the x-axis of the tensor. 
This representation proves to be particularly valuable for analyzing the symmetry-constrained magnetoelectric response in crystalline materials, establishing a clear connection between polar lattice dynamics and magnetic properties.
In addition, the plotted vector field is color-mapped ( Fig.~\ref{fig:dmc_visual}.d) ) by converting each vector's coordinates into RGB values.
This directional color-coded technique enhances the perception of the vector field's structure allowing for immediate visual identification of atoms with similar/opposite vector orientations.

Additionally, the resulting vector field can be decomposed into irreducible tensor representations \cite{Urru2022,Verbeek2023,Spaldin2013} that directly correspond to magnetoelectric multipoles: a scalar trace component corresponds to the magnetic monopole contribution, a fully antisymmetric tensor corresponds to the toroidal moment, and a symmetric traceless tensor corresponds to the magnetic quadrupole term (see Appendix~\ref{Appendix:Multipoles} for the mathematical formalism and Fig.~\ref{fig:F2DmcDecomposition} for a visual aspect). 
Each of these magnetoelectric multipoles represents distinct physical mechanisms for magnetoelectric coupling, with the monopole $a$ term relating to isotropic effects, the quadrupole $q$ capturing anisotropic interactions, and the toroidal moment $t$ describing vortex-like response of the magnetization to electric field stimuli.
Fig.~\ref{fig:F2DmcDecomposition} illustrates this decomposition for the basal Fluorine atom at the 8h Wyckoff site. 
This analysis disentangles the complex total response into its irreducible symmetry-adapted components, revealing that the magnetoelectric properties at this site are governed by a combination of toroidal and quadrupolar moments. 
Visually, the toroidal contribution manifests as a distinct curl (vortex) field, indicating a rotational character in the coupling between lattice displacement and induced magnetization. 
Conversely, the quadrupolar contribution appears as a shear field, reflecting the anisotropic nature of the response.

The DMC magnetization texture generated by the individual Wyckoff sites constitutes a continuous global vector field that strictly obeys the crystal symmetry. 
This continuity implies that the symmetry-allowed magnetization at any point in the unit cell, including unoccupied Wyckoff positions, must naturally align with the overall ``flow'' of the vector field established by the occupied sites. 
This coherence confirms that the derived DMC tensors provide a consistent and robust description of the magnetic symmetry throughout the crystal volume.

Figure \ref{fig:dmc_visual} provides a vectorial representation of the dynamical magnetic charge tensors for the occupied Wyckoff sites. In these plots, the vector field describes the local magnetoelectric response: the arrows indicate the direction and relative magnitude of the magnetization induced by an atomic displacement along the corresponding radial direction. 
Due to the reciprocal nature of the coupling, this visualization can equivalently be interpreted as the displacement direction of the atom by an external magnetic field. 
A particularly high-symmetry case is observed for the apical Fluorine at Wyckoff site 4b; the associated vector field is quasi purely divergent, physically, this indicates that the induced magnetization is mostly parallel to the atomic displacement vector regardless of the direction of motion ($\mathbf{M} \parallel \mathbf{u}$), a signature a monopolar response.

The calculation of the DMC in the cubic phase is not possible using our ``frozen-phonon'' technique.
Indeed, moving atoms one by one breaks the symmetry that allows the first-order Jahn-Teller electronic instability to relax for some cases, with the consequence that it is not anymore in the linear response regime. 

\subsection{Anti-magnetoelectric response in KCoF$_3$}

The magnetic space group of the ground state phase of KCoF$_3$ with the magnetic moment along the $z$ direction is $I4'/mcm'$,  does not allow for a magnetoelectric response.
However, as the DMCs are non-zero, the contribution to the antimagnetoelectric response stemming from the spin-lattice, respectively orbital-lattice couplings is non-zero.
Assisted by the visualization tool for examining atomic responses to external fields of Fig.~\ref{fig:dmc_visual}, it is more apparent why polar modes do not induce a magnetization.
Indeed, atoms occupying equivalent Wyckoff positions with opposite DMC (like in KCoF$_3$) and moving in the same direction with the same amplitude will necessarily cancel each other regarding the induced magnetization, and resulting in an anti-magnetoelectric behavior.

By splitting the response through magnetic sublattice, we have found that each sublattice carries a total response of 105 ps/m per formula unit (f.u.), i.e. 210 ps/m per spin-channel (the unit cell contains 2 f.u. with spin up and two f.u. with spin down), which is split between the spin-lattice $\alpha^{\boldsymbol{S},latt} = 50$ ps/m per f.u. and the orbital-lattice $\alpha^{\boldsymbol{L},latt} = 55$ ps/m per f.u. contributions.
Hence, the local equivalent magnetoelectric response is rather large.


\section{Discussion}
To check how general is the possibility to have non zero DMCs and to have anti-magnetoelectric, we have looked from symmetry analysis how they are in the most common $Pnma$ perovskite phase within the ferromagnetic, C-, A- and G-type AFM magnetic orders.
We report these results in Tab. S.V of the Supplementary Material.
Although none of those phases are magnetoelectric, we can see that all of the possible magnetic space groups that can be built from the $Pnma$ perovskites are anti-magnetoelectric as they all have one or more of the Wyckoff sites that host a non-zero DMC tensor.
The DMC are strictly forbidden in the paramagnetic parent group ($Pnma1'$), i.e. when there is no magnetic order.
The same result is observed if one does the same general analysis in both Jahn-Teller phases from $R$ and $M$ zone boundary points or with octahedra rotation in one, two or three directions (not shown).
Hence, we can see that the anti-magnetoelectric property is actually very common in perovskites even for those that  are non-functional.


\section{Conclusions}
\label{sec:Conclusions}

In summary, we have studied from first-principles DFT calculations the cubic to tetragonal phase transition of KCoF$_3$ by looking at the phonons, energy landscape with respect to atom displacements and the magnetic ground state and the associated DMC.
We could confirm that the phase transition is from a first-order Jahn-Teller electronic instability and that the associated ground state phase is anti-magnetoelectric.
We found that the DMCs of Co are zero and that the ones of K and F atoms are very large, the largest value being reached by the apical fluorine anion when moving it along the x or y direction perpendicular to the tetragonal axe.
All of the large values are evenly split between spin and orbital magnetization, which sum to make particularly large values.
We could demonstrate from symmetry arguments the shape of the DMC tensors of each atom and that the DMC of Co cation should indeed be zero, even though it is the source of magnetism in KCoF$_3$ crystal.

The presence of a Jahn-Teller orbital ordering seems to be an important condition to obtain large DMC; however; this would require a more systematic and deeper analysis, which is beyond the scope of the present paper.
We can however discuss here the possible reasons of why the DMCs are large in KCoF$_3$.
The strong correlation between the Jahn-Teller distortion and the emergence of a large DMC can be understood through the intimate coupling of electronic orbitals, lattice degrees of freedom, and spin-orbit interactions. The Jahn-Teller  effect is, by definition, a manifestation of strong electron-phonon coupling, where a structural distortion lifts the electronic degeneracy of the magnetic ion's $d$-orbitals (in this case, Co$^{2+}$) to lower the system's total energy. 
This distortion not only modifies the crystal field environment but also partially quenches the orbital angular momentum. Consequently, the orbital component of the magnetic moment becomes highly sensitive to the positions of the surrounding ligands (F$^{-}$ ions). 
A dynamical atomic displacement, which corresponds to a phonon excitation, modulates the Jahn-Teller  distortion pattern. 
This dynamic modulation of the local crystal field, in turn, induces a change in the orbital magnetization.
Through spin-orbit coupling, this orbital fluctuation is directly coupled to the spin degrees of freedom, inducing a corresponding change in the spin magnetization.
The DMC is thus significantly enhanced because the Jahn-Teller  mechanism could be a direct and efficient pathway for a lattice vibration to dynamically control both the orbital and spin magnetic moments.
This would require testing this hypothesis in several compounds with and without Jahn-Teller distortions to verify it.

We could also show from symmetry arguments that the anti-magnetoelectric response is present in most of the non-functional phases of perovskites and that it could be a more general property of magnetic materials that goes beyond the scope of multiferroic or magnetoelectric crystals.
We hope that our work will encourage further studies and use of the DMC in  magnetic materials.


\section*{Acknowledgements}

The authors acknowledge the FNRS and the Excellence of Science program (EOS ``ShapeME'', No. 40007525) funded by the FWO and F.R.S.-FNRS. Computational resources have been provided by the Consortium des \'Equipements de Calcul Intensif (C\'ECI), funded by the Fonds de la Recherche Scientifique (F.R.S.-FNRS) under Grant No. 2.5020.11 and the Tier-1 Lucia supercomputer of the Walloon Region, infrastructure funded by the Walloon Region under the grant agreement n°1910247 and by the High Performance Computing Mesocenter of the University of Lille financed by the University, the Hauts-de-France Region, the State, the FEDER and the University's laboratories through a pooling process.

\appendix
\section{Dynamical magnetic charges (DMC)}
\label{annex:DMC}


First, let us reassess the importance of the spin and orbital magnetic moments coupled to an electric field.
The linear magnetoelectric effect, which is the coupling between the electric and magnetic degrees of freedom in materials, can be simply traced from the net magnetization \textit{$\mathbf{M}$} (polarization $\mathbf{P}$) induced by the applied electric (magnetic) field's amplitude:

\begin{equation}
    \alpha_{ij}=\mu_{0}\frac{\delta M_{j}}{\delta \mathcal{E}_{i}}\biggr\rvert_{\mathcal{H}} =\frac{\delta P_{i}}{\delta \mathcal{H}_{j}}\biggr\rvert_{\mathcal{E}}.
    \label{eq:alpha}
\end{equation}
From the theoretical standpoint, the second-rank tensor $\alpha_{ij}$ can be decomposed into three contributions: electronic, lattice and strain, $\alpha^{tot} = \alpha^{\text{elec}} + \alpha^{\text{latt}} + \alpha^{\text{str}}$, each of which arising from the coupling with either the spin, $\boldsymbol{S}$, or the orbital, $\boldsymbol{L}$, degrees of freedom. While the electronic component ($\alpha^{\text{elec}}$) response requires the presence of a spontaneous polarization to be nonzero, it is not the case for the latter two. In fact, lattice-mediated terms can give rise to a significant response in the vicinity of a ferroelectric phase transition \cite{Bousquet2011a, Iniguez2008}. Furthermore, the full magnetoelectric response can be strongly enhanced as a results of an unquenched orbital magnetic moment due to a strong lattice-orbital coupling \cite{Malashevich2012, Scaramucci2012,Solovyev2016}.
We remind from a previous work~\cite{Braun2024,Iniguez2008} that the lattice-mediated contribution to the magnetoelectric tensor takes the following form:
\begin{equation}
    \alpha^{\text{latt}}_{ij} = \frac{\mu_{0}}{\Omega} \sum_{n=1}^{N_{\text{IR}}} \frac{ S_{n,ij} }{\omega_{n}^2},
    \label{eq:alpha_latt_1}
\end{equation}
\noindent
where $\omega_n$ the phonon \textit{angular} frequencies corresponding to the $N_{\text{IR}}$ infrared-active phonon modes, and  $S_{n,ij}$ is the magnetoelectric mode-oscillator strength tensor, defined as follows:
\begin{equation}
S_{n,ij}=\left(\sum_{\kappa,i'}Z^{*\text{m}}_{\kappa,ii'}{u}_{n,\kappa,i'}\right)\times\left(\sum_{\kappa,j'}Z^{*\text{e}}_{\kappa,j'j}{u}_{n,\kappa,j'}\right),
    \label{ME_Mode-oscillator}
\end{equation}
where $\kappa$ identifies the atom in the unit cell,  ${u}_{n,\kappa,j}$ the normalized eigendisplacement, $Z_\kappa^{*\text{e}}$ and $Z_\kappa^{*\text{m}}$  the Born effective charge (BEC) and dynamical magnetic charge (DMC), respectively, defined as:
\begin{equation}
    \label{eq:becs}
    Z^{*\text{e}}_{\kappa,ij} = \Omega \frac{\partial P_{i}}{\partial \tau_{\kappa,j}} =\frac{\partial^2E}{\partial\mathcal{E}_{i}\partial \tau_{\kappa,j}},
\end{equation}
\noindent 
i.e., the net polarization along direction $i$ induced by a displacement $\tau$ of atom $\kappa$ along direction $j$, and
\begin{equation}
    \label{eq:dmcs}
    Z^{*\text{m}}_{\kappa,ij} = \Omega \frac{\partial M_{i}}{\partial \tau_{\kappa,j}}=\frac{\partial^2E}{\partial\mathcal{H}_{i}\partial \tau_{\kappa,j}},
\end{equation}
\noindent 
which corresponds to the net magnetization along the direction $i$, induced by a displacement $\tau$ of atom $\kappa$ along direction $j$~\cite{Iniguez2008,Ye2014}, or equivalently in the DFPT framework, the change in energy of a system under an applied magnetic field $\mathcal{H}$ along direction $i$ and a displacement $\tau$ of atom $\kappa$ along direction $j$.

Those DMCs have been used recently to explain the anti-magnetoelectric case~\cite{Verbeek2023} where the global magnetoelectric response is zero but the local one is not, similarly to antiferromagnetism with  non zero atom magnetic moments that are anti-aligned from site to site and giving  zero total magnetization~\cite{Braun2024}.

While DMCs were defined through magnetoelectric, magnetodielectric and multiferroic materials~\cite{Iniguez2008,Ye2014,Kimura2016,Braun2024}, in this work we show that large DMC can exists in non-functional materials. 
We study the anti-magnetoelectric effect in the antiferromagnetically ordered KCoF$_3$ perovskite crystal in the presence of a Jahn-Teller instability by means of \textit{ab~initio} calculations.

\section{Magnetoelectric Multipoles}\label{Appendix:Multipoles}

Magnetoelectric (ME) multipoles emerge as fundamental descriptors of magnetic structures lacking combined space-inversion (\(P\)) and time-reversal (\(T\)) symmetry, i.e., \(PT\)-non-invariant systems~\cite{Thole2018, Yatsushiro2021Multipole, Bhowal2022Hidden}. They originate from the spatial inhomogeneity of the magnetization density \(\mathbf{M}(\mathbf{r})\) within a magnetic unit cell, formally arising from the first spatial moment tensor of magnetization, \(\mathcal{M}_{ij} = \int r_i M_j(\mathbf{r}) \, d^3r\), where the integral is typically taken over the unit cell~\cite{Thole2018, Yatsushiro2021Multipole}. This tensor \(\mathcal{M}_{ij}\) inherently captures the \(PT\)-odd character because the position vector \(\mathbf{r}\) is \(P\)-odd, \(T\)-even, while the magnetization \(\mathbf{M}\) is \(P\)-even, \(T\)-odd, making their product \(r_i M_j\) odd under both \(P\) and \(T\) transformations. Consequently, ME multipoles are intrinsically linked to linear magnetoelectric phenomena, characterized by the ME tensor \(\alpha_{ij}\)~\cite{Thole2018}. The microscopic ME multipoles defined via \(\mathcal{M}_{ij}\) provide a direct connection to the macroscopic response tensor \(\alpha_{ij}\).

The tensor \(\mathcal{M}_{ij}\) can be decomposed into its irreducible components, corresponding to distinct ME multipoles~\cite{Thole2018, Yatsushiro2021Multipole}:
\begin{itemize}
\item(i) The ME monopole (also related to the anapole), associated with the trace of \(\mathcal{M}\), \(\text{Tr}(\mathcal{M}) = \int \mathbf{r} \cdot \mathbf{M}(\mathbf{r}) \, d^3r\), . It represents a point-like source of \(PT\)-odd characteristics.
\item(ii) The magnetic toroidal moment \(\mathbf{T}\), an axial vector derived from the antisymmetric part of \(\mathcal{M}_{ij}\): \(T_k = \frac{1}{2} \epsilon_{ijk} \mathcal{M}_{ij} = \frac{1}{2} \int (\mathbf{r} \times \mathbf{M}(\mathbf{r}))_k \, d^3r\). Physically, it corresponds to vortex-like arrangements of magnetic moments, breaking both \(P\) and \(T\) symmetries individually~\cite{Thole2018, Yatsushiro2021Multipole}. It couples to the curl of an electric field or the time derivative of a magnetic field.
\item(iii) The ME quadrupole tensor \(Q^{\text{ME}}_{ij}\),obtained from the symmetric traceless part of \(\mathcal{M}_{ij}\): \(Q^{\text{ME}}_{ij} = \frac{1}{2} \int (r_i M_j + r_j M_i - \frac{2}{3} \delta_{ij} \mathbf{r} \cdot \mathbf{M}) \, d^3r\).
\end{itemize}
The magnetoelectric multipoles collectively reconstitute the tensor $\mathcal{M}_{ij}$ shown in Eq.~\ref{eq_me_tensor}. This tensor directly corresponds to the magnetoelectric tensor $\alpha_{ij}$ when constrained by the magnetic point group symmetry of the unit cell, which determines both the allowed Wyckoff positions and their corresponding magnetic moment orientations. Furthermore, $\mathcal{M}_{ij}$ represents the dynamical magnetic charge when further restricted by the local symmetries at specific Wyckoff sites.

\begin{equation}
\label{eq_me_tensor}
\resizebox{\linewidth}{!}{$
\mathcal{M}_{ij}=
\begin{pmatrix}
a + \frac{1}{2}q_{x^2-y^2} - \frac{1}{2} q_{z^2} 
& t_z + q_{xy} & t_y + q_{xz}\\
-t_z + q_{xy} & a - \frac{1}{2}q_{x^2-y^2} - \frac{1}{2} q_{z^2}  & -t_x + q_{yz} \\
-t_y + q_{xz} & t_x + q_{yz} &  a + q_{z^2}
\end{pmatrix}
$}
\end{equation}

While initially explored primarily in non-centrosymmetric magnetic insulators, the significance of ME multipoles is increasingly recognized even in metallic systems where itinerant electrons might be expected to screen electric fields~\cite{Thole2018, Yatsushiro2021Multipole} and in non magnetoelectric systems. They serve as crucial primary or secondary order parameters for complex magnetic phases, particularly "hidden orders" where the net magnetic dipole moment vanishes but a non-zero ME multipole exists~\cite{Thole2018, Yatsushiro2021Multipole, Bhowal2021Revealing, Bhowal2022Hidden}. Their presence can distinguish between magnetic structures that are degenerate at the dipole level but differ in their ME multipolar content i.e Fe$_{2}$O$_{3}$ and Cr$_{2}$O$_{3}$\cite{Verbeek2023}, leading to distinct macroscopic properties like non-reciprocal transport, optical effects (e.g., directional dichroism), and specific contributions to magnetoelectric responses~\cite{Yatsushiro2021Multipole, Thole2018}. Theoretically, their identification and calculation often rely on symmetry analysis (group theory) and first-principles electronic structure methods~\cite{Thole2018, Bhowal2021Revealing}. Experimentally, detecting ME multipoles can be challenging as their signatures are often weaker than those of magnetic dipoles. Techniques sensitive to \(PT\)-symmetry breaking, such as optical second harmonic generation (SHG), non-linear optical effects, resonant and non-resonant X-ray diffraction (exploiting specific symmetries or dichroism), Compton scattering~\cite{Bhowal2021Revealing}, and detailed analysis of polarized neutron scattering data (going beyond the standard magnetic structure factor), are employed for their investigation~\cite{Bhowal2022Hidden, Yatsushiro2021Multipole}. 
To the best of our knowledge, there is no experimental measurement of DMC and/or ME multipoles.

\end{document}


\title{Supplementary Material: First-principles study of KCoF$_3$: Jahn-Teller effect, dynamical magnetic charges, magnetoelectric multipoles and antimagnetoelectricity}

\author{Bogdan Guster}
\affiliation{Physique Th\'eorique des Mat\'eriaux, QMAT, CESAM, Universit\'e de Li\`ege, B-4000 Sart-Tilman, Belgium}
\affiliation{European Theoretical Spectroscopy Facility, www.etsf.eu}
\author{Maxime Braun}
\affiliation{Physique Th\'eorique des Mat\'eriaux, QMAT, CESAM, Universit\'e de Li\`ege, B-4000 Sart-Tilman, Belgium}
\affiliation{Univ. Lille, CNRS, Centrale Lille, ENSCL, Univ. Artois, UMR 8181-UCCS-Unité de Catalyse et Chimie du Solide, F-59000 Lille, France}
\author{Eric Bousquet}
\affiliation{Physique Th\'eorique des Mat\'eriaux, QMAT, CESAM, Universit\'e de Li\`ege, B-4000 Sart-Tilman, Belgium}

\maketitle

\section{Supplementary Tables}


\begin{table*}[!hptb]
\caption{Relative energy difference ($\Delta E$) between the cubic and the Jahn-Teller and the elastic phase with respect to the Hubbard $U$ value. The corresponding lattice parameters of each case is listed as well. All lengths are expressed in Bohr.}

\begin{tabular}{|c|ccc|cccc|cccc|}
\hline
\#                    & \multicolumn{3}{c|}{Cubic}                                                                                            & \multicolumn{4}{c|}{JT - Q$^{2-}$}                                                                                                                                             & \multicolumn{4}{c|}{Monoclinic}                                                                                        \\ \hline
\multirow{2}{*}{U}    & \multicolumn{1}{c|}{\multirow{2}{*}{a}}      & \multicolumn{1}{c|}{\multirow{2}{*}{b}}      & \multirow{2}{*}{c}      & \multicolumn{1}{c|}{\multirow{2}{*}{a}}      & \multicolumn{1}{c|}{\multirow{2}{*}{b}}      & \multicolumn{1}{c|}{\multirow{2}{*}{c}}      & \multirow{2}{*}{$\Delta E$ (meV/f.u.)} & \multicolumn{1}{c|}{a}      & \multicolumn{1}{c|}{b}      & \multicolumn{1}{c|}{c}      & \multirow{2}{*}{$\Delta E$ (meV/f.u.)} \\ \cline{9-11}
                      & \multicolumn{1}{c|}{}                        & \multicolumn{1}{c|}{}                        &                         & \multicolumn{1}{c|}{}                        & \multicolumn{1}{c|}{}                        & \multicolumn{1}{c|}{}                        &                        & \multicolumn{1}{c|}{$\alpha$}  & \multicolumn{1}{c|}{$\beta$}   & \multicolumn{1}{c|}{$\gamma$}   &                        \\ \hline
\multirow{2}{*}{4.0}  & \multicolumn{1}{c|}{\multirow{2}{*}{10.695}} & \multicolumn{1}{c|}{\multirow{2}{*}{10.695}} & \multirow{2}{*}{15.658} & \multicolumn{1}{c|}{\multirow{2}{*}{10.880}} & \multicolumn{1}{c|}{\multirow{2}{*}{10.880}} & \multicolumn{1}{c|}{\multirow{2}{*}{15.124}} & \multirow{2}{*}{-23.1} & \multicolumn{1}{c|}{10.885} & \multicolumn{1}{c|}{10.885} & \multicolumn{1}{c|}{15.123} & \multirow{2}{*}{-0.7}  \\ \cline{9-11}
                      & \multicolumn{1}{c|}{}                        & \multicolumn{1}{c|}{}                        &                         & \multicolumn{1}{c|}{}                        & \multicolumn{1}{c|}{}                        & \multicolumn{1}{c|}{}                        &                        & \multicolumn{1}{c|}{90}     & \multicolumn{1}{c|}{90}     & \multicolumn{1}{c|}{88.00}  &                        \\ \hline
\multirow{2}{*}{4.5}  & \multicolumn{1}{c|}{\multirow{2}{*}{10.696}} & \multicolumn{1}{c|}{\multirow{2}{*}{10.696}} & \multirow{2}{*}{15.658} & \multicolumn{1}{c|}{\multirow{2}{*}{10.887}} & \multicolumn{1}{c|}{\multirow{2}{*}{10.887}} & \multicolumn{1}{c|}{\multirow{2}{*}{15.137}} & \multirow{2}{*}{-22.1} & \multicolumn{1}{c|}{10.887} & \multicolumn{1}{c|}{10.892} & \multicolumn{1}{c|}{15.133} & \multirow{2}{*}{-0.7}  \\ \cline{9-11}
                      & \multicolumn{1}{c|}{}                        & \multicolumn{1}{c|}{}                        &                         & \multicolumn{1}{c|}{}                        & \multicolumn{1}{c|}{}                        & \multicolumn{1}{c|}{}                        &                        & \multicolumn{1}{c|}{90}     & \multicolumn{1}{c|}{90}     & \multicolumn{1}{c|}{88.03}  &                        \\ \hline
\multirow{2}{*}{5.0}  & \multicolumn{1}{c|}{\multirow{2}{*}{10.709}} & \multicolumn{1}{c|}{\multirow{2}{*}{10.709}} & \multirow{2}{*}{15.658} & \multicolumn{1}{c|}{\multirow{2}{*}{10.890}} & \multicolumn{1}{c|}{\multirow{2}{*}{10.890}} & \multicolumn{1}{c|}{\multirow{2}{*}{15.130}} & \multirow{2}{*}{-21.1} & \multicolumn{1}{c|}{10.893} & \multicolumn{1}{c|}{10.897} & \multicolumn{1}{c|}{15.137} & \multirow{2}{*}{-0.7}  \\ \cline{9-11}
                      & \multicolumn{1}{c|}{}                        & \multicolumn{1}{c|}{}                        &                         & \multicolumn{1}{c|}{}                        & \multicolumn{1}{c|}{}                        & \multicolumn{1}{c|}{}                        &                        & \multicolumn{1}{c|}{90}     & \multicolumn{1}{c|}{90}     & \multicolumn{1}{c|}{88.05}  &                        \\ \hline
\multirow{2}{*}{6.0}  & \multicolumn{1}{c|}{\multirow{2}{*}{10.719}} & \multicolumn{1}{c|}{\multirow{2}{*}{10.719}} & \multirow{2}{*}{15.663} & \multicolumn{1}{c|}{\multirow{2}{*}{10.898}} & \multicolumn{1}{c|}{\multirow{2}{*}{10.898}} & \multicolumn{1}{c|}{\multirow{2}{*}{15.158}} & \multirow{2}{*}{-19.5} & \multicolumn{1}{c|}{10.899} & \multicolumn{1}{c|}{10.901} & \multicolumn{1}{c|}{15.163} & \multirow{2}{*}{-0.7}  \\ \cline{9-11}
                      & \multicolumn{1}{c|}{}                        & \multicolumn{1}{c|}{}                        &                         & \multicolumn{1}{c|}{}                        & \multicolumn{1}{c|}{}                        & \multicolumn{1}{c|}{}                        &                        & \multicolumn{1}{c|}{90}     & \multicolumn{1}{c|}{90}     & \multicolumn{1}{c|}{88.15}  &                        \\ \hline
\multirow{2}{*}{8.0}  & \multicolumn{1}{c|}{\multirow{2}{*}{10.727}} & \multicolumn{1}{c|}{\multirow{2}{*}{10.727}} & \multirow{2}{*}{15.667} & \multicolumn{1}{c|}{\multirow{2}{*}{10.910}} & \multicolumn{1}{c|}{\multirow{2}{*}{10.910}} & \multicolumn{1}{c|}{\multirow{2}{*}{15.206}} & \multirow{2}{*}{-17.2} & \multicolumn{1}{c|}{10.915} & \multicolumn{1}{c|}{10.911} & \multicolumn{1}{c|}{15.191} & \multirow{2}{*}{-1.1}  \\ \cline{9-11}
                      & \multicolumn{1}{c|}{}                        & \multicolumn{1}{c|}{}                        &                         & \multicolumn{1}{c|}{}                        & \multicolumn{1}{c|}{}                        & \multicolumn{1}{c|}{}                        &                        & \multicolumn{1}{c|}{90}     & \multicolumn{1}{c|}{90}     & \multicolumn{1}{c|}{88.21}  &                        \\ \hline
\multirow{2}{*}{10.0} & \multicolumn{1}{c|}{\multirow{2}{*}{10.765}} & \multicolumn{1}{c|}{\multirow{2}{*}{10.765}} & \multirow{2}{*}{15.677} & \multicolumn{1}{c|}{\multirow{2}{*}{10.923}} & \multicolumn{1}{c|}{\multirow{2}{*}{10.923}} & \multicolumn{1}{c|}{\multirow{2}{*}{15.232}} & \multirow{2}{*}{-14.8} & \multicolumn{1}{c|}{10.926} & \multicolumn{1}{c|}{10.928} & \multicolumn{1}{c|}{15.224} & \multirow{2}{*}{-0.7}  \\ \cline{9-11}
                      & \multicolumn{1}{c|}{}                        & \multicolumn{1}{c|}{}                        &                         & \multicolumn{1}{c|}{}                        & \multicolumn{1}{c|}{}                        & \multicolumn{1}{c|}{}                        &                        & \multicolumn{1}{c|}{90}     & \multicolumn{1}{c|}{90}     & \multicolumn{1}{c|}{88.33}  &                        \\ \hline
\end{tabular}
\end{table*}

\begin{table}[!hptb]
\caption{Calculated Born effective charges tensor components (in unit of the elemental charge $e$) in the FM phase of KCoF$_3$ in the cubic geometry. Columns refer to derivatives with respect to the electric field and lines to derivatives with respect to atomic positions, in atomic units.}
\begin{tabular}{c|crrr}
\hline\hline
Atom                  & \multicolumn{4}{c}{Born Effective Charge (e)} \\ \hline\hline
                      &          & \multicolumn{1}{c}{$\partial/\partial \mathcal{E}_{x}$}         & \multicolumn{1}{c}{$\partial/\partial \mathcal{E}_{y}$}        & \multicolumn{1}{c}{$\partial/\partial \mathcal{E}_{z}$}        \\
\multirow{3}{*}{K}    & $\partial/\partial \tau_{x}$       & 1.19       & 0         & 0         \\
                      & $\partial/\partial \tau_{y}$       & 0          & 1.19      & 0         \\
                      & $\partial/\partial \tau_{z}$       & 0          & 0         & 1.21      \\ \hline
Co                    & $\partial/\partial \tau_{x}$       & 2.01       & 0.        & 0         \\
                      & $\partial/\partial \tau_{y}$       & 0          & 2.01      & 0         \\
                      & $\partial/\partial \tau_{z}$       & 0          & 0         & 2.39      \\ \hline
\multirow{3}{*}{F$_{api}$} & $\partial/\partial \tau_{x}$  & -0.82     & 0         & 0         \\
                      & $\partial/\partial \tau_{y}$       & 0          & -0.82     & 0         \\
                      & $\partial/\partial \tau_{z}$       & 0          & 0         & -1.64     \\ \hline
\multirow{3}{*}{F$_{bas}$} & $\partial/\partial \tau_{x}$  & -0.63      & 0.        & 0         \\
                      & $\partial/\partial \tau_{y}$       & 0          & -1.67     & 0         \\
                      & $\partial/\partial \tau_{z}$       & 0          & 0         & -0.82     \\ \hline
\end{tabular}
\label{tab:cubic_becs}
\end{table}
\begin{table}[!hptb]
\caption{Calculated Born effective charges tensor components (in unit of the elemental charge $e$) in the G-AFM phase of KCoF$_3$ in the R-point Jahn-Teller distored geometry. Columns refer to derivatives with respect to the electric field and lines to derivatives with respect to atomic positions, in atomic units.}
\begin{tabular}{c|crrr}
\hline\hline
Atom                  & \multicolumn{4}{c}{Born Effective Charge (e)} \\ \hline\hline
                      &          & \multicolumn{1}{c}{$\partial/\partial \mathcal{E}_{x}$}         & \multicolumn{1}{c}{$\partial/\partial \mathcal{E}_{y}$}        & \multicolumn{1}{c}{$\partial/\partial \mathcal{E}_{z}$}        \\
\multirow{3}{*}{K}    & $\partial/\partial \tau_{x}$       & 1.20       & 0         & 0         \\
                      & $\partial/\partial \tau_{y}$       & 0          & 1.20      & 0         \\
                      & $\partial/\partial \tau_{z}$       & 0          & 0         & 1.21      \\ \hline
Co                    & $\partial/\partial \tau_{x}$       & 2.13       & -0.18     & 0         \\
                      & $\partial/\partial \tau_{y}$       & -0.18      & 2.13      & 0         \\
                      & $\partial/\partial \tau_{z}$       & 0          & 0         & 2.31      \\ \hline
\multirow{3}{*}{F$_{api}$} & $\partial/\partial \tau_{x}$       & -0.78      & 0         & 0         \\
                      & $\partial/\partial \tau_{y}$       & 0          & -0.78     & 0         \\
                      & $\partial/\partial \tau_{z}$       & 0          & 0         & -1.84     \\ \hline
\multirow{3}{*}{F$_{bas}$} & $\partial/\partial \tau_{x}$       & -1.26      & 0.49      & 0         \\
                      & $\partial/\partial \tau_{y}$       & 0.49       & -1.26     & 0         \\
                      & $\partial/\partial \tau_{z}$      & 0          & 0         & -0.83     \\ \hline
\end{tabular}
\label{tab:jt_becs}
\end{table}

\begin{table}[!hptb]
\caption{Mode effective charges of the polar phonon modes of KCoF$_3$ in the JT phase.}
\begin{tabular}{c|c|ccc}
\hline\hline
 \multirow{2}{*}{$\omega$(meV)}                       & \multirow{2}{*}{Character} & \multicolumn{3}{c}{Mode effective charge}                       \\ \cline{3-5} 
 \multicolumn{1}{c|}{}                                &                            & \multicolumn{1}{c|}{x}         & \multicolumn{1}{c|}{y}         & \multicolumn{1}{c}{z}  \\ \hline
 15                                                   & A$_{2u}$                   & \multicolumn{1}{r|}{0}         & \multicolumn{1}{r|}{0}         & 2.78                   \\ \hline
 15                                                   & E$_u$                      & \multicolumn{1}{r|}{0}         & \multicolumn{1}{r|}{-2.79}     & 0                      \\ \hline
 15                                                   & E$_u$                      & \multicolumn{1}{r|}{2.79}      & \multicolumn{1}{r|}{0}         &                        \\ \hline
 21                                                   & E$_u$                      & \multicolumn{1}{r|}{0}         & \multicolumn{1}{r|}{-0.11}     & 0                      \\ \hline
 21                                                   & E$_u$                      & \multicolumn{1}{r|}{-0.11}     & \multicolumn{1}{r|}{0}         & 0                      \\ \hline
 26                                                   & A$_{2u}$                   & \multicolumn{1}{r|}{0}         & \multicolumn{1}{r|}{0}         & -3.29                  \\ \hline
 27                                                   & E$_u$                      & \multicolumn{1}{r|}{0}         & \multicolumn{1}{r|}{-2.93}     & 0                      \\ \hline
 27                                                   & E$_u$                      & \multicolumn{1}{r|}{-2.93}     & \multicolumn{1}{r|}{0}         & 0                      \\ \hline
 33                                                   & E$_u$                      & \multicolumn{1}{r|}{-0.94}     & \multicolumn{1}{r|}{0}         & 0                      \\ \hline
 33                                                   & E$_u$                      & \multicolumn{1}{r|}{0}         & \multicolumn{1}{r|}{0.94}      & 0                      \\ \hline
 54                                                   & E$_u$                      & \multicolumn{1}{r|}{-0.001}    & \multicolumn{1}{r|}{4.081}     & 0                      \\ \hline
 54                                                   & E$_u$                      & \multicolumn{1}{r|}{-4.081}    & \multicolumn{1}{r|}{-0.001}    & 0                      \\ \hline
 56                                                   & A$_{2u}$                   & \multicolumn{1}{r|}{0}         & \multicolumn{1}{r|}{0}         & -4.27                  \\ \hline\hline
\end{tabular}
\end{table}
\newcommand{\tensorial}[1]{%
  \ensuremath{\left[ \begin{smallmatrix} #1 \end{smallmatrix} \right]}%
}

\begin{longtable}{clllc}
\caption{Magnetic space groups associated with the parent space group $Pnma$ (\#62). The table lists the BNS number and Wyckoff positions filled in standard perovskites ($ABX_3$). \label{tab:pnma_magnetic}} \\
\toprule
\textbf{Group Name} & \textbf{Wyckoff} & \textbf{Representative Site} & \textbf{Site} & \textbf{Tensor Shape} \\
\textbf{(BNS \#)} & \textbf{Pos.} & \textbf{$(x,y,z \mid m_x,m_y,m_z)$} & \textbf{Sym.} & \textbf{$[Z^{*m}_{ij}]$} \\
\midrule
\endfirsthead

\multicolumn{5}{c}%
{{\bfseries \tablename\ \thetable{} -- continued from previous page}} \\
\toprule
\textbf{Group Name} & \textbf{Pos.} & \textbf{Representative Site} & \textbf{Sym.} & \textbf{Tensor Shape} \\
\midrule
\endhead

\midrule
\multicolumn{5}{r}{{Continued on next page}} \\
\bottomrule
\endfoot

\bottomrule
\endlastfoot

\multirow{3}{*}[-0.5em]{\shortstack{$Pnma1'$\\(\#62.442)}} 
 & $4b$ & $(0,0,1/2 \mid 0,0,0)$ & $-11'$ & $\varnothing$ \\
 & $4c$ & $(x,1/4,z \mid 0,0,0)$ & $.m.1'$ & $\varnothing$ \\
 & $8d$ & $(x,y,z \mid 0,0,0)$ & $11'$ & $\varnothing$ \\
\midrule

\multirow{3}{*}[-2em]{\shortstack{$Pn'ma$\\(\#62.443)}} 
 & $4b$ & $(0,0,1/2 \mid 0,0,0)$ & $-1'$ & \tensorial{Z^{*m}_{xx}&Z^{*m}_{xy}&Z^{*m}_{xz}\\Z^{*m}_{yx}&Z^{*m}_{yy}&Z^{*m}_{yz}\\Z^{*m}_{zx}&Z^{*m}_{zy}&Z^{*m}_{zz}} \\
 & $4c$ & $(x,1/4,z \mid 0,m_y,0)$ & $.m.$ & \tensorial{0&Z^{*m}_{xy}&0\\Z^{*m}_{yx}&0&Z^{*m}_{yz}\\0&Z^{*m}_{zy}&0} \\
 & $8d$ & $(x,y,z \mid m_x,m_y,m_z)$ & $1$ & \tensorial{Z^{*m}_{xx}&Z^{*m}_{xy}&Z^{*m}_{xz}\\Z^{*m}_{yx}&Z^{*m}_{yy}&Z^{*m}_{yz}\\Z^{*m}_{zx}&Z^{*m}_{zy}&Z^{*m}_{zz}} \\
\midrule

\multirow{3}{*}[-2em]{\shortstack{$Pnm'a$\\(\#62.444)}} 
 & $4b$ & $(0,0,1/2 \mid 0,0,0)$ & $-1'$ & \tensorial{Z^{*m}_{xx}&Z^{*m}_{xy}&Z^{*m}_{xz}\\Z^{*m}_{yx}&Z^{*m}_{yy}&Z^{*m}_{yz}\\Z^{*m}_{zx}&Z^{*m}_{zy}&Z^{*m}_{zz}} \\
 & $4c$ & $(x,1/4,z \mid m_x,0,m_z)$ & $.m'.$ & \tensorial{Z^{*m}_{xx}&0&Z^{*m}_{xz}\\0&Z^{*m}_{yy}&0\\Z^{*m}_{zx}&0&Z^{*m}_{zz}} \\
 & $8d$ & $(x,y,z \mid m_x,m_y,m_z)$ & $1$ & \tensorial{Z^{*m}_{xx}&Z^{*m}_{xy}&Z^{*m}_{xz}\\Z^{*m}_{yx}&Z^{*m}_{yy}&Z^{*m}_{yz}\\Z^{*m}_{zx}&Z^{*m}_{zy}&Z^{*m}_{zz}} \\
\midrule

\multirow{3}{*}[-2em]{\shortstack{$Pnma'$\\(\#62.445)}} 
 & $4b$ & $(0,0,1/2 \mid 0,0,0)$ & $-1'$ & \tensorial{Z^{*m}_{xx}&Z^{*m}_{xy}&Z^{*m}_{xz}\\Z^{*m}_{yx}&Z^{*m}_{yy}&Z^{*m}_{yz}\\Z^{*m}_{zx}&Z^{*m}_{zy}&Z^{*m}_{zz}} \\
 & $4c$ & $(x,1/4,z \mid 0,m_y,0)$ & $.m.$ & \tensorial{0&Z^{*m}_{xy}&0\\Z^{*m}_{yx}&0&Z^{*m}_{yz}\\0&Z^{*m}_{zy}&0} \\
 & $8d$ & $(x,y,z \mid m_x,m_y,m_z)$ & $1$ & \tensorial{Z^{*m}_{xx}&Z^{*m}_{xy}&Z^{*m}_{xz}\\Z^{*m}_{yx}&Z^{*m}_{yy}&Z^{*m}_{yz}\\Z^{*m}_{zx}&Z^{*m}_{zy}&Z^{*m}_{zz}} \\
\midrule

\multirow{3}{*}[-1.5em]{\shortstack{$Pn'm'a$\\(\#62.446)}} 
 & $4b$ & $(0,0,1/2 \mid m_x,m_y,m_z)$ & $-1$ & $\varnothing$ \\
 & $4c$ & $(x,1/4,z \mid m_x,0,m_z)$ & $.m'.$ & \tensorial{Z^{*m}_{xx}&0&Z^{*m}_{xz}\\0&Z^{*m}_{yy}&0\\Z^{*m}_{zx}&0&Z^{*m}_{zz}} \\
 & $8d$ & $(x,y,z \mid m_x,m_y,m_z)$ & $1$ & \tensorial{Z^{*m}_{xx}&Z^{*m}_{xy}&Z^{*m}_{xz}\\Z^{*m}_{yx}&Z^{*m}_{yy}&Z^{*m}_{yz}\\Z^{*m}_{zx}&Z^{*m}_{zy}&Z^{*m}_{zz}} \\
\midrule

\multirow{3}{*}[-1.5em]{\shortstack{$Pnm'a'$\\(\#62.447)}} 
 & $4b$ & $(0,0,1/2 \mid m_x,m_y,m_z)$ & $-1$ & $\varnothing$ \\
 & $4c$ & $(x,1/4,z \mid m_x,0,m_z)$ & $.m'.$ & \tensorial{Z^{*m}_{xx}&0&Z^{*m}_{xz}\\0&Z^{*m}_{yy}&0\\Z^{*m}_{zx}&0&Z^{*m}_{zz}} \\
 & $8d$ & $(x,y,z \mid m_x,m_y,m_z)$ & $1$ & \tensorial{Z^{*m}_{xx}&Z^{*m}_{xy}&Z^{*m}_{xz}\\Z^{*m}_{yx}&Z^{*m}_{yy}&Z^{*m}_{yz}\\Z^{*m}_{zx}&Z^{*m}_{zy}&Z^{*m}_{zz}} \\
\midrule

\multirow{3}{*}[-1.5em]{\shortstack{$Pn'ma'$\\(\#62.448)}} 
 & $4b$ & $(0,0,1/2 \mid m_x,m_y,m_z)$ & $-1$ & $\varnothing$ \\
 & $4c$ & $(x,1/4,z \mid 0,m_y,0)$ & $.m.$ & \tensorial{0&Z^{*m}_{xy}&0\\Z^{*m}_{yx}&0&Z^{*m}_{yz}\\0&Z^{*m}_{zy}&0} \\
 & $8d$ & $(x,y,z \mid m_x,m_y,m_z)$ & $1$ & \tensorial{Z^{*m}_{xx}&Z^{*m}_{xy}&Z^{*m}_{xz}\\Z^{*m}_{yx}&Z^{*m}_{yy}&Z^{*m}_{yz}\\Z^{*m}_{zx}&Z^{*m}_{zy}&Z^{*m}_{zz}} \\
\midrule

\multirow{3}{*}[-2em]{\shortstack{$Pn'm'a'$\\(\#62.449)}} 
 & $4b$ & $(0,0,1/2 \mid 0,0,0)$ & $-1'$ & \tensorial{Z^{*m}_{xx}&Z^{*m}_{xy}&Z^{*m}_{xz}\\Z^{*m}_{yx}&Z^{*m}_{yy}&Z^{*m}_{yz}\\Z^{*m}_{zx}&Z^{*m}_{zy}&Z^{*m}_{zz}} \\
 & $4c$ & $(x,1/4,z \mid m_x,0,m_z)$ & $.m'.$ & \tensorial{Z^{*m}_{xx}&0&Z^{*m}_{xz}\\0&Z^{*m}_{yy}&0\\Z^{*m}_{zx}&0&Z^{*m}_{zz}} \\
 & $8d$ & $(x,y,z \mid m_x,m_y,m_z)$ & $1$ & \tensorial{Z^{*m}_{xx}&Z^{*m}_{xy}&Z^{*m}_{xz}\\Z^{*m}_{yx}&Z^{*m}_{yy}&Z^{*m}_{yz}\\Z^{*m}_{zx}&Z^{*m}_{zy}&Z^{*m}_{zz}} \\
\midrule

\multirow{4}{*}[-2.5em]{\shortstack{$P_anma$\\(\#62.450)}} 
 & $8c$ & $(3/4,1/2,1/2 \mid 0,0,0)$ & $-1'$ & \tensorial{Z^{*m}_{xx}&Z^{*m}_{xy}&Z^{*m}_{xz}\\Z^{*m}_{yx}&Z^{*m}_{yy}&Z^{*m}_{yz}\\Z^{*m}_{zx}&Z^{*m}_{zy}&Z^{*m}_{zz}} \\
 & $8d$ & $(0,1/2,1/2 \mid m_x,m_y,m_z)$ & $-1$ & $\varnothing$ \\
 & $8f$ & $(x,1/4,z \mid 0,m_y,0)$ & $.m.$ & \tensorial{0&Z^{*m}_{xy}&0\\Z^{*m}_{yx}&0&Z^{*m}_{yz}\\0&Z^{*m}_{zy}&0} \\
 & $16g$ & $(x,y,z \mid m_x,m_y,m_z)$ & $1$ & \tensorial{Z^{*m}_{xx}&Z^{*m}_{xy}&Z^{*m}_{xz}\\Z^{*m}_{yx}&Z^{*m}_{yy}&Z^{*m}_{yz}\\Z^{*m}_{zx}&Z^{*m}_{zy}&Z^{*m}_{zz}} \\
\midrule

\multirow{4}{*}[-2em]{\shortstack{$P_bnma$\\(\#62.451)}} 
 & $4b$ & $(0,0,0 \mid m_x,0,m_z)$ & $.2'/m'.$ & $\varnothing$ \\
 & $4d$ & $(0,0,1/2 \mid m_x,0,m_z)$ & $.2'/m'.$ & $\varnothing$ \\
 & $8g$ & $(x,1/4,z \mid 0,m_y,0)$ & $.m.$ & \tensorial{0&Z^{*m}_{xy}&0\\Z^{*m}_{yx}&0&Z^{*m}_{yz}\\0&Z^{*m}_{zy}&0} \\
 & $16i$ & $(x,y,z \mid m_x,m_y,m_z)$ & $1$ & \tensorial{Z^{*m}_{xx}&Z^{*m}_{xy}&Z^{*m}_{xz}\\Z^{*m}_{yx}&Z^{*m}_{yy}&Z^{*m}_{yz}\\Z^{*m}_{zx}&Z^{*m}_{zy}&Z^{*m}_{zz}} \\
\midrule

\multirow{0}{*}[-0em]{\shortstack{$P_cnma$\\(\#62.452)}} 
 & $8a$ & $(0,0,3/4 \mid 0,0,0)$ & $-1'$ & \tensorial{Z^{*m}_{xx}&Z^{*m}_{xy}&Z^{*m}_{xz}\\Z^{*m}_{yx}&Z^{*m}_{yy}&Z^{*m}_{yz}\\Z^{*m}_{zx}&Z^{*m}_{zy}&Z^{*m}_{zz}} \\
 & $8b$ & $(0,0,0 \mid m_x,m_y,m_z)$ & $-1$ & $\varnothing$ \\
 & $8d$ & $(x,1/4,z \mid 0,m_y,0)$ & $.m.$ & \tensorial{0&Z^{*m}_{xy}&0\\Z^{*m}_{yx}&0&Z^{*m}_{yz}\\0&Z^{*m}_{zy}&0} \\
 & $16e$ & $(x,y,z \mid m_x,m_y,m_z)$ & $1$ & \tensorial{Z^{*m}_{xx}&Z^{*m}_{xy}&Z^{*m}_{xz}\\Z^{*m}_{yx}&Z^{*m}_{yy}&Z^{*m}_{yz}\\Z^{*m}_{zx}&Z^{*m}_{zy}&Z^{*m}_{zz}} \\
\midrule

\multirow{3}{*}[-1.5em]{\shortstack{$P_Anma$\\(\#62.453)}} 
 & $8d$ & $(0,0,0 \mid m_x,m_y,m_z)$ & $-1$ & $\varnothing$ \\
 & $8f$ & $(x,3/4,z \mid 0,m_y,0)$ & $.m.$ & \tensorial{0&Z^{*m}_{xy}&0\\Z^{*m}_{yx}&0&Z^{*m}_{yz}\\0&Z^{*m}_{zy}&0} \\
 & $16h$ & $(x,y,z \mid m_x,m_y,m_z)$ & $1$ & \tensorial{Z^{*m}_{xx}&Z^{*m}_{xy}&Z^{*m}_{xz}\\Z^{*m}_{yx}&Z^{*m}_{yy}&Z^{*m}_{yz}\\Z^{*m}_{zx}&Z^{*m}_{zy}&Z^{*m}_{zz}} \\
\midrule

\multirow{3}{*}[-2em]{\shortstack{$P_Bnma$\\(\#62.454)}} 
 & $8d$ & $(1/4,0,1/4 \mid 0,0,0)$ & $-1'$ & \tensorial{Z^{*m}_{xx}&Z^{*m}_{xy}&Z^{*m}_{xz}\\Z^{*m}_{yx}&Z^{*m}_{yy}&Z^{*m}_{yz}\\Z^{*m}_{zx}&Z^{*m}_{zy}&Z^{*m}_{zz}} \\
 & $8g$ & $(x,1/4,z \mid 0,m_y,0)$ & $.m.$ & \tensorial{0&Z^{*m}_{xy}&0\\Z^{*m}_{yx}&0&Z^{*m}_{yz}\\0&Z^{*m}_{zy}&0} \\
 & $16h$ & $(x,y,z \mid m_x,m_y,m_z)$ & $1$ & \tensorial{Z^{*m}_{xx}&Z^{*m}_{xy}&Z^{*m}_{xz}\\Z^{*m}_{yx}&Z^{*m}_{yy}&Z^{*m}_{yz}\\Z^{*m}_{zx}&Z^{*m}_{zy}&Z^{*m}_{zz}} \\
\midrule

\multirow{3}{*}[-1.5em]{\shortstack{$P_Cnma$\\(\#62.455)}} 
 & $8c$ & $(0,0,0 \mid m_x,m_y,m_z)$ & $-1$ & $\varnothing$ \\
 & $8f$ & $(x,1/4,z \mid 0,m_y,0)$ & $.m.$ & \tensorial{0&Z^{*m}_{xy}&0\\Z^{*m}_{yx}&0&Z^{*m}_{yz}\\0&Z^{*m}_{zy}&0} \\
 & $16g$ & $(x,y,z \mid m_x,m_y,m_z)$ & $1$ & \tensorial{Z^{*m}_{xx}&Z^{*m}_{xy}&Z^{*m}_{xz}\\Z^{*m}_{yx}&Z^{*m}_{yy}&Z^{*m}_{yz}\\Z^{*m}_{zx}&Z^{*m}_{zy}&Z^{*m}_{zz}} \\
\midrule

\multirow{4}{*}[-2em]{\shortstack{$P_Inma$\\(\#62.456)}} 
 & $4a$ & $(0,0,0 \mid 0,m_y,m_z)$ & $2'/m'..$ & $\varnothing$ \\
 & $4b$ & $(0,0,1/2 \mid 0,m_y,m_z)$ & $2'/m'..$ & $\varnothing$ \\
 & $8i$ & $(x,1/4,z \mid 0,m_y,0)$ & $.m.$ & \tensorial{0&Z^{*m}_{xy}&0\\Z^{*m}_{yx}&0&Z^{*m}_{yz}\\0&Z^{*m}_{zy}&0} \\
 & $16j$ & $(x,y,z \mid m_x,m_y,m_z)$ & $1$ & \tensorial{Z^{*m}_{xx}&Z^{*m}_{xy}&Z^{*m}_{xz}\\Z^{*m}_{yx}&Z^{*m}_{yy}&Z^{*m}_{yz}\\Z^{*m}_{zx}&Z^{*m}_{zy}&Z^{*m}_{zz}} \\

\end{longtable}

\begin{longtable}{clllc}
\caption{Wyckoff Positions, Site Symmetries, and DMC shapes for Space Group \#140.545. Only Wyckoff positions filled in standard perovskites are listed.} \\
\toprule
\textbf{Group Name} & \textbf{Wyckoff} & \textbf{Representative Site} & \textbf{Site} & \textbf{Tensor Shape} \\
\textbf{(BNS \#)} & \textbf{Pos.} & \textbf{$(x,y,z \mid m_x,m_y,m_z)$} & \textbf{Sym.} & \textbf{$[Z^{*m}_{ij}]$} \\
\midrule
\endfirsthead

\multicolumn{5}{c}%
{{\bfseries \tablename\ \thetable{} -- continued from previous page}} \\
\toprule
\textbf{Group Name} & \textbf{Pos.} & \textbf{Representative Site} & \textbf{Sym.} & \textbf{Tensor Shape} \\
\midrule
\endhead

\midrule
\multicolumn{5}{r}{{Continued on next page}} \\
\bottomrule
\endfoot

\bottomrule
\endlastfoot

\multirow{4}{*}[-2.5em]{\shortstack{$I4'/mcm'$\\(\#140.545)}} 
 & $4a$ & $(0,0,1/4 \mid 0,0,0)$ & $4'22'$ & \tensorial{-Z^{*m}_{xx}&0&0\\0&Z^{*m}_{xx}&0\\0&0&0} \\
 & $4b$ & $(0,1/2,1/4 \mid 0,0,0)$ & $-4'2m'$ & \tensorial{Z^{*m}_{xx}&0&0\\0&Z^{*m}_{xx}&0\\0&0&Z^{*m}_{zz}} \\
 & $4d$ & $(0,0,0 \mid 0,0,mz)$ & $m.m'm'$ & $\varnothing$  \\
 & $8h$ & $(x,x+1/2,0 \mid 0,0,m_z)$ & $m.2'm'$ & \tensorial{0&0&Z^{*m}_{xz}\\0&0&Z^{*m}_{xz}\\Z^{*m}_{zx}&Z^{*m}_{zx}&0} \\

\end{longtable}

\clearpage
\newpage

\section{Supplementary Figures}



\begin{figure}[!hptb]
    \centering
    \includegraphics[width=0.5\linewidth]{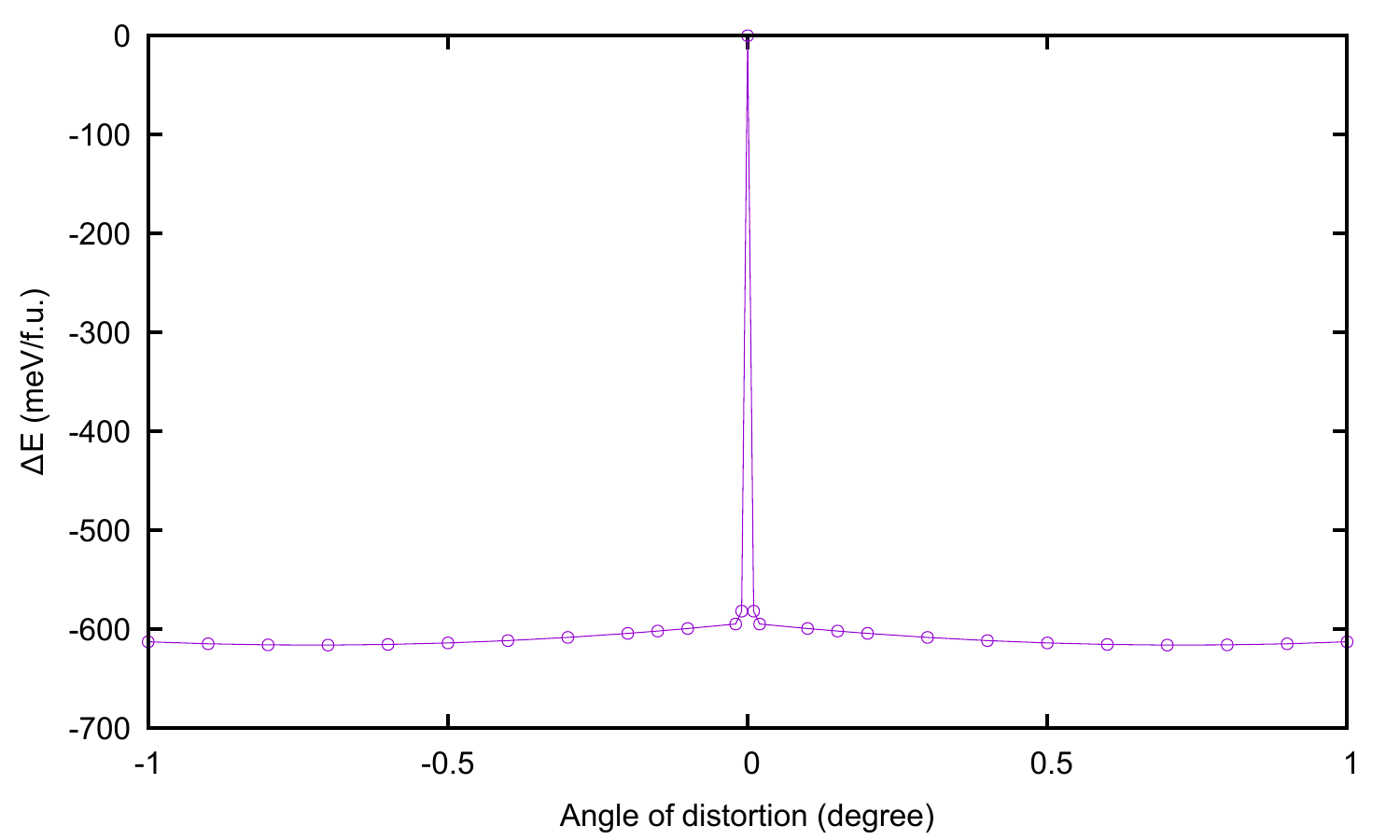}
    \caption{Energy evolution under the first-order type Jahn-Teller rigid distortion in KCoF$_3$ in the \textit{collinear treatment} of the G-AFM magnetic state. The energy gain is comparable with the work of Varignon et al. \cite{Varignon2019} (See Fig. 7). }
    \label{fig:delta_energy_collinear}
\end{figure}

\begin{figure}[!hptb]
    \centering
    \includegraphics[width=0.5\linewidth]{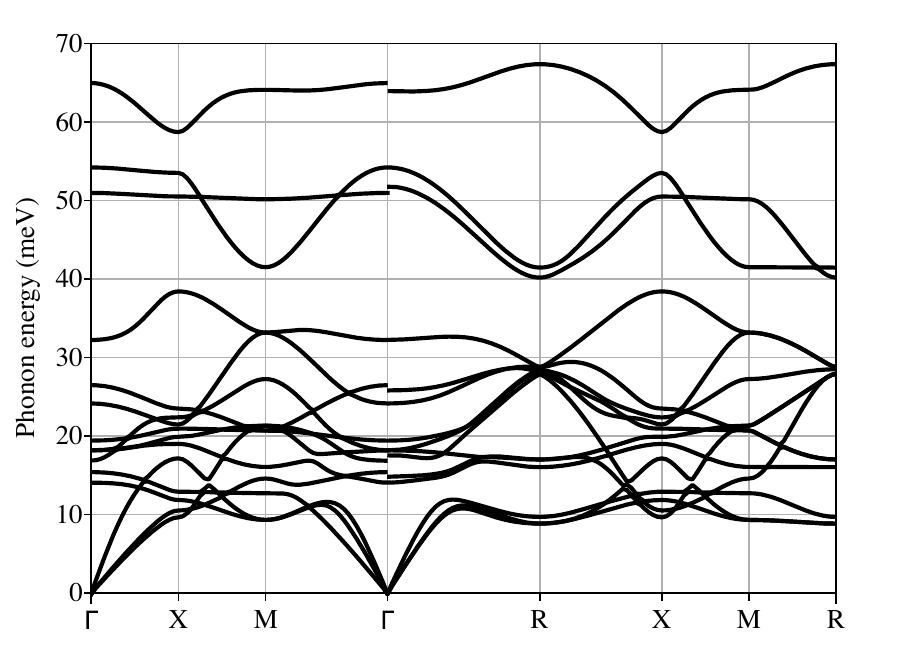}
    \caption{Phonon dispersion of KCoF$_3$ in the collinear treatment of the FM undistorted phase. This is in line with the G-AFM phonon dispersion calculation of Dubrovin et al.~\cite{Dubrovin2021} where they also observe no unstable phonon mode.}
    \label{fig:phonon_jt}
\end{figure}


%

%